\documentclass[12pt]{article}

\usepackage{newtxtext,newtxmath}

\usepackage{graphicx}

\usepackage[letterpaper,margin=1in]{geometry}

\renewenvironment{abstract}
	{\quotation}
	{\endquotation}

\date{}

\makeatletter
\renewcommand{\fnum@figure}{\textbf{Figure \thefigure}}
\renewcommand{\fnum@table}{\textbf{Table \thetable}}
\makeatother

\usepackage{scicite}

\usepackage{url}

\def\scititle{
	Melting upon cooling in a quantum magnet
}
\title{\bfseries \boldmath \scititle}

\author{
	K.~Jakseti\v c$^{1,2}$,
	T.~Arh$^{1,2,3}$,
	M.~Pregelj$^{1,2}$,
    M.~Gomil\v{s}ek$^{1,2}$,
    M.~Dragomir$^{1}$,\\ 
    P.~Prelov\v{s}ek$^{1}$, 
    M.~Ulaga$^{4}$,
    L.~\v{S}ibav$^{1}$, 
    M.~Malovrh$^{2}$, 
    K.~\v{Z}eleznikar$^{2}$, 
    Z.~Jagli\v{c}i\'c$^{5,6}$,\\ 
    P.~Manuel$^{7}$,
    F.~Orlandi$^{7}$, 
    D.~Khalyavin$^{7}$, 
    M.~D. Le$^{7}$,
    N.~Bujault$^{8}$, 
    E.~Lhotel$^{8}$,\\
    J.~van~Tol$^{9}$, 
    U.~Jena$^{10}$, 
    B.~Sana$^{10}$, 
    P.~Khuntia$^{10,11\ast}$, 
    and A.~Zorko$^{1,2\ast}$\and
	\small$^{1}$Jo\v{z}ef Stefan Institute, Jamova cesta~39, 1000 Ljubljana, Slovenia.\\
	\small$^{2}$Faculty of Mathematics and Physics, University of Ljubljana, Jadranska ulica~19, 1000 Ljubljana, Slovenia.\\
    \small$^{3}$PSI Center for Neutron and Muon Sciences CNM, CH-5232 Villigen PSI, Switzerland.\\
    \small$^{4}$Max Planck Institute for Physics of Complex Systems, Dresden, Germany.\\
    \small$^{5}$Faculty of Civil and Geodetic Engineering, University of Ljubljana, 1000 Ljubljana, Slovenia.\\
    \small$^{6}$Institute of Mathematics, Physics and Mechanics, 1000 Ljubljana, Slovenia.\\
    \small$^{7}$ISIS facility, Rutherford Appleton Laboratory, Chilton, Didcot, OX11 0QX, Oxfordshire, UK.\\
    \small$^{8}$Institut N\'eel, CNRS \& Universit\'e Grenoble Alpes, 38000 Grenoble, France.\\
    \small$^{9}$National High Magnetic Field Laboratory, Florida State University, Tallahassee, Florida 32310, USA.\\
    \small$^{10}$Department of Physics, Indian Institute of Technology Madras, Chennai 600 036, India.\\
    \small$^{11}$Quantum Centre of Excellence for Diamond and Emergent Materials, Indian Institute of Technology\\
    \small Madras, Chennai, 600036, India.\and
	\small$^\ast$Corresponding author. Email: pkhuntia@iitm.ac.in \\
    \small$^{\ast}$Corresponding author. Email: andrej.zorko@ijs.si
}

\begin{document} 
\maketitle
\newpage
\begin{abstract} \bfseries \boldmath
Heating enhances thermal fluctuations and typically leads to melting of solids, but 
in exceptional cases, heating can also cause liquids to solidify.
The paradigm of this counterintuitive phenomenon is solidification of liquid $^3$He upon increasing temperature, known as the Pomeranchuk effect.
Here we show that such inverse melting also appears in quantum magnetism. 
We find that, on cooling, the Ising-like triangular-lattice antiferromagnet erbium heptatantalate first develops a three-sublattice long-range magnetic order---analogous to a solid---which then, unexpectedly, melts at even lower temperatures into a short-range correlated spin-stripe state---analogous to a liquid.
We propose that such an unprecedented ``spin Pomeranchuk effect" can generically arise from strong competition between spin-spin interactions in frustrated magnets, and provides a novel avenue to transformations between exotic magnetic phases.
\end{abstract}

\noindent

Atoms generally develop long-range positional order, i.e., they crystallize below a freezing temperature when thermal fluctuations become sufficiently weak.
Although rare, there are exceptions to this universal trend. 
A famous example is liquid helium, which resists solidification at low pressure even at absolute zero temperature due to strong quantum zero-point motion of atoms~\cite{beamish2020mechanical}.
The $^3$He nevertheless solidifies under moderate applied pressure, but surprisingly, it does not transforms from liquid to solid upon cooling, but rather upon heating, which is known as the Pomeranchuk effect~\cite{pomeranchuk1950theory,lee1997extraordinary}.
Here, the solid phase of $^3$He is stabilized at high enough temperatures due to excess entropy arising from disordered nuclear spins, lowering its free energy below that of the liquid phase.
Although scarce, this inverse-melting phenomenon has been lately observed in diverse contexts beyond crystalline ordering, from vortex states in high-temperature superconductors~\cite{avraham2001inverse} to domain textures in ferromagnetic~\cite{portmann2003inverse} and ferroelectric thin films~\cite{nahas2020inverse}, and to correlated states in magic-angle twisted bilayer graphene~\cite{rozen2021entropic, saito2021isospin}.
As dictated by thermodynamics, in all these cases the high-temperature ordered phase possesses larger total entropy than the low-temperature disordered phase.
The establishment of a finite order parameter and the associated loss of the configurational entropy at elevated temperatures must, therefore, be accompanied by release of some other microscopic degrees of freedom~\cite{schupper2005inverse}; e.g., nuclear spins become nearly independent in solid $^3$He~\cite{pomeranchuk1950theory,lee1997extraordinary}, impurity-pinning disorder gets enhanced due to vortex crystallization~\cite{avraham2001inverse}, topological disorder intensifies in ordered textures of thin films~\cite{nahas2020inverse,schupper2005inverse}, and magnetic degrees of freedom become active in the temperature-stabilized  hybrid correlated state of the magic-angle twisted bilayer graphene~\cite{rozen2021entropic}.  
A Pomeranchuk-like transition has also been predicted in magnetism, but remarkably, here, the involvement of any additional degrees of freedom to spin is not required~\cite{hassan2007supersolidity, schupper2004spin}.

Magnetic analogues of liquid helium are short-range correlated spin states, for instance, highly entangled quantum spin liquids (QSL)~\cite{balents2010spin,savary2017quantum,broholm2020quantum,lancaster2023quantum}. 
In these, long-range orientational ordering of spins that is usually established in magnets below a transition temperature, is absent even at zero temperature due to strong quantum fluctuations.
Such fluctuations typically arise from competing spin--spin interactions on frustrated spin lattices. 
At the same time, frustration can amplify entropic effects, even in longe-range ordered magnetic phases~\cite{hassan2007supersolidity}.
As a result, ordered (solid) magnetic phases might be stabilized at elevated temperatures, as long as they possess more entropy than the disordered strongly correlated (liquid) ground state.
This would lead to unprecedented magnetic ordering upon heating, analogous to the Pomeranchuk effect.
\begin{figure}[h!]
\centering
\includegraphics[trim = 0mm 0mm 0mm 0mm, clip, width=0.85\linewidth]{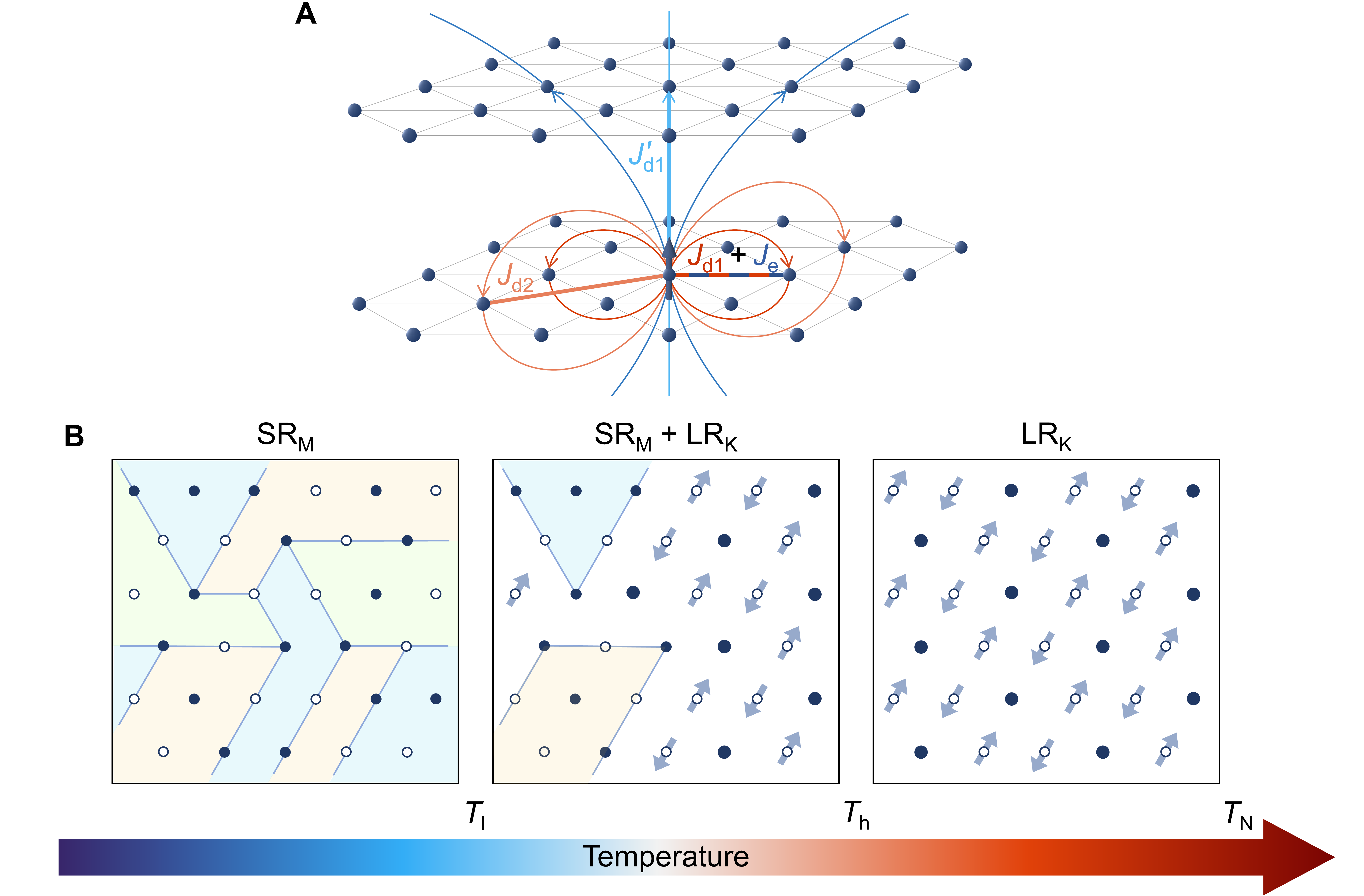}
\caption{\textbf{Magnetic interactions and correlated phases of ErTa$_7$O$_{19}$.}
(\textbf{A}) 
Perfect triangular planes of Er$^{3+}$ magnetic moments. 
The effective spins 1/2 of the ground-state Er$^{3+}$ Kramers doublets are coupled via XXZ antiferromagnetic nearest-neighbor exchange interaction $J_\text{e}$, and Ising-like antiferromagnetic intra-plane $J_{\text{d}i}$ and ferromagnetic inter-plane $J'_{\text{d}i}$ long-range dipolar interactions with neighbors $i$.
(\textbf{B}) 
The temperature evolution of magnetic phases, where $\text{SR}_\text{M}$ denotes the low-temperature 
phase with short-range spin-stripe-type correlations, corresponding to the magnetic propagation vector ${\bf q}_\text{M}=(1/2,0,0)$ (M points in the Brillouin zone), and $\text{LR}_\text{K}$ is the three-sublattice long-range ordered spin supersolid phase with magnetic propagation vector ${\bf q}_\text{K}=(1/3,1/3,0)$ (K points). 
The $\text{SR}_\text{M}$ and $\text{LR}_\text{K}$ phases coexist in a broad temperature range between $T_\text{l} \simeq 25$~mK and $T_\text{h} = 65$~mK, while the latter remains stable up to the N\'eel temperature $T_\text{N} = 100$~mK. 
Open and full circles correspond to the up (U) and down (D) out-of-plane spin components, respectively, while the arrows show the orientation of the in-plane spin components in the spin supersolid.
Solid blue lines highlight the boundaries of the small, color-coded $\text{SR}_\text{M}$ domains.
}
\label{fig1}
\end{figure}

Here we demonstrate the first experimental realization of the Pomeranchuk effect in quantum magnetism---the ``spin Pomeranchuk effect" that appears in the triangular-lattice Ising-like antiferromagnet ErTa$_7$O$_{19}$ (Fig.~\ref{fig1}A). 
This compound is a new member of the rare-earth (RE) heptatantalate family that has recently been introduced as a promising platform for realization of QSL ground states~\cite{arh2022ising}.
Such a state was, indeed, found in NdTa$_7$O$_{19}$~\cite{arh2022ising}, and later also suggested for other members of the family~\cite{li2025possible, baiwa2025quantum}. 
In stark contrast, our experiments reveal that ErTa$_7$O$_{19}$ undergoes a phase transition to a long-range ordered three-sublattice state ($\text{LR}_\text{K}$) below $T_\text{N} = 100$~mK, with the magnetic propagation vector ${\bf q}_\text{K}=(1/3,1/3,0)$ corresponding to the K points of the Brillouin zone (Fig.~\ref{fig1}B).
Furthermore, this state is characterized by large excess entropy.
These characteristics are typical for a spin supersolid---a state that simultaneously breaks translational and rotational symmetry~\cite{wang2009extended, heidarian2010supersolidity, yamamoto2014quantum, sellmann2015phase, gao2022spin}---where excess entropy at finite temperatures originates from enhanced spin fluctuations~\cite{xiang2024giant}.
When interactions extend beyond nearest neighbors, the spin supersolid may no longer be the lowest-energy state~\cite{gallegos2025phase}, undermining its stability at low temperatures. 
Indeed, in ErTa$_7$O$_{19}$, we observe that the solid $\text{LR}_\text{K}$ state gradually transforms into a short-range correlated spin-stripe state ($\text{SR}_\text{M}$) between $T_\text{h} = 65$~mK and $T_\text{l} \simeq 25$~mK (Fig.~\ref{fig1}B).
The spin correlations in the $\text{SR}_\text{M}$ state occur at the M points of the Brillouin zone, corresponding to the magnetic propagation vector ${\bf q}_\text{M}=(1/2,0,0)$, but, surprisingly, extend across only a few unit cells.
The transformation from this liquid-like $\text{SR}_\text{M}$ state to the solid-like $\text{LR}_\text{K}$ state upon heating is directly analogous to the Pomeranchuk effect in $^3$He.
However, in contrast to other known realization of this phenomenon, it does not require the involvement of multiple degrees of freedom.


\begin{figure}[h!]
\centering
\includegraphics[trim = 0mm 0mm 0mm 0mm, clip, width=1\linewidth]{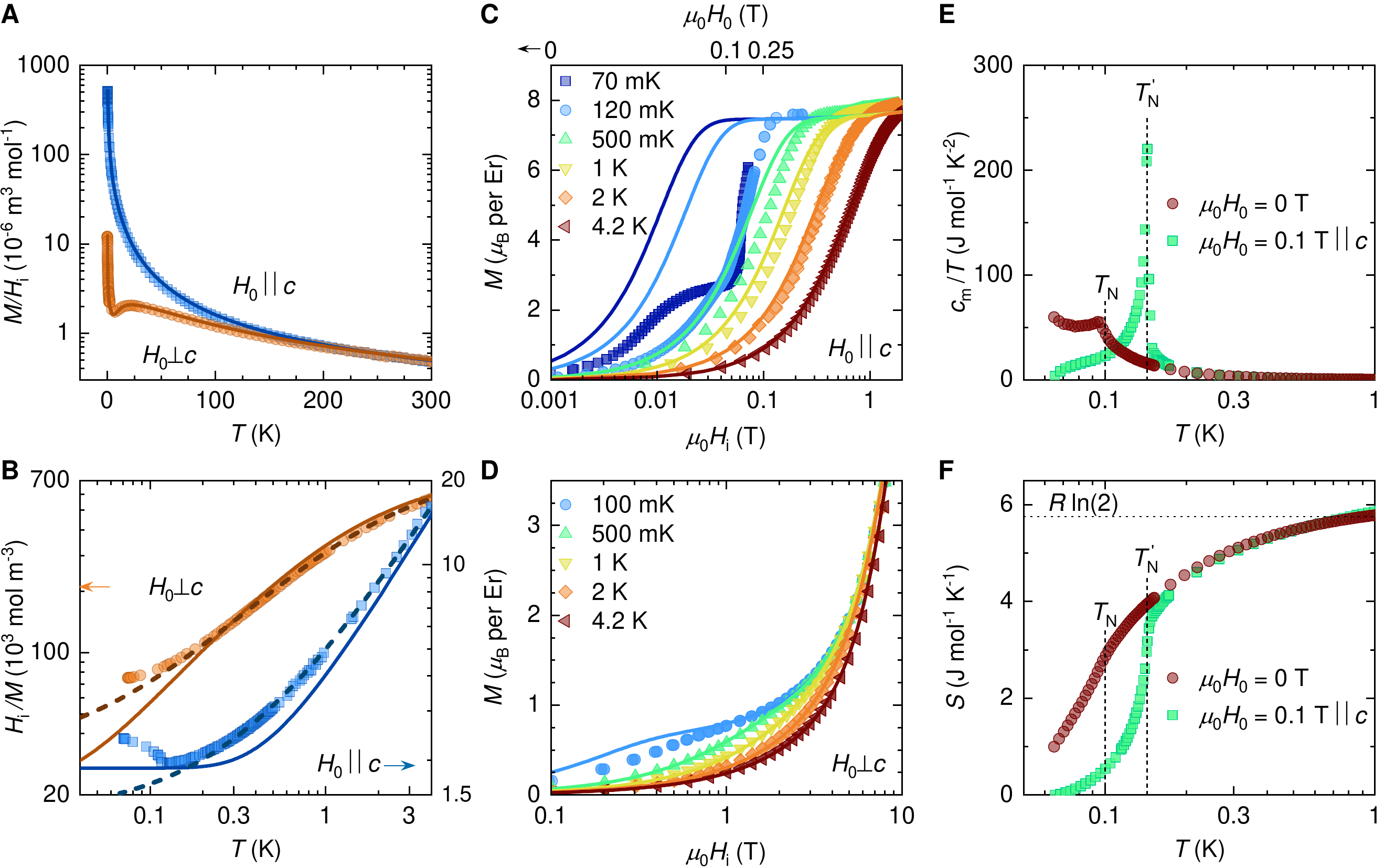}
\caption{{\bf Bulk properties of ErTa$_7$O$_{19}$.}
(\textbf{A})
Normalized magnetization $M/H_\text{i}$ for an external field $\mu_0 H_0 = 0.1$~T applied either perpendicular ($H_0{\parallel} c$) or within ($H_0{\perp} c$) the triangular planes. 
$M$ is magnetization, $H_\text{i} = H_0 - NM$ the internal magnetic field, and $N$ demagnetization factor~\cite{methods}.
The solid lines are fits with the crystal-electric-field (CEF) model~\cite{methods}.
(\textbf{B}) 
The inverse normalized magnetization at low temperatures compared to the CEF model (solid lines) and fitted with a Weiss molecular-field approximation of the spin--spin interactions (dashed lines)~\cite{methods}. 
(\textbf{C,D})
Isothermal magnetization curves for $H_0 {\parallel} c$ and $H_0{\perp} c$ versus $H_\text{i}$ at several temperatures. 
Solid lines show the fits to the CEF model.
The upper axis in panel \textbf{C} corresponds to the applied field $H_0$ for the 70-mK dataset, which shows a 1/3-plateau phase at intermediate fields (see also figure~\ref{figS6}c).
(\textbf{E}) 
The temperature dependence of the magnetic specific heat divided by temperature under zero field and in a field $\mu_0 H_0 = 0.1$~T applied along $c$, which induces the 1/3-plateau phase at low temperatures.
(\textbf{F}) 
The magnetic entropy versus temperature.
The dotted horizontal line is the full entropy of a spin-1/2 system, $R\,\rm ln$(2), where $R$ is the gas constant.
The dashed lines in panels \textbf{E} and \textbf{F} show the N\'{e}el temperature $T_\text{N} = 100$~mK under zero applied field and the transition temperature $T_\text{N}' = 143$~mK into the 1/3-plateau phase in a 0.1~T field applied along $c$.}
\label{fig2}
\end{figure}

\subsection*{Anisotropic magnetism}
Like other RE heptatantalates, ErTa$_7$O$_{19}$ possesses perfect triangular-lattice symmetry~\cite{methods, sibav2025optimized}. 
At temperatures below 150~K, strong magnetic anisotropy develops (Fig.~\ref{fig2}A) with an easy axis perpendicular to the triangular planes (i.e., along the crystallographic $c$ axis). 
At low temperatures, the isothermal magnetization saturates at $\sim$8$\mu_\text{B}$ per magnetic ion already in sub-tesla fields $H_0{\parallel} c$ (Fig.~\ref{fig2}C), while no such saturation is observed up to 8~T when $H_0{\perp} c$ (Fig.~\ref{fig2}D).
Such anisotropic magnetic behavior is a consequence of the crystal-electric-field (CEF), splitting the $^4I_{15/2}$ spin-orbit multiplet of the Er$^{3+}$ ion into Kramers doublets.
The transitions from the CEF ground-state doublet into excited doublets result in flat modes observed in our inelastic neutron scattering (INS) experiment (figure~\ref{figS2}A--C).
Using a CEF model\cite{methods}, we can accurately reproduce the positions and intensities of these modes (figure~\ref{figS2}D--F) together with the temperature dependence of anisotropic normalized magnetization (Fig.~\ref{fig2}A) and isothermal magnetization above $\sim$2~K (Fig.~\ref{fig2}C,D).
The modeling reveals the composition and anisotropic $g$ factors of all 8 Er$^{3+}$ Kramers doublets (table~\ref{CEF-vec}).
The ground-state doublet is a pure dipole state with principal $g$ values $g_0^\text{c}=14.9\pm 0.2$ and $g_0^\text{ab}=1.2\pm 0.2$, as directly confirmed by electron spin resonance measurements~\cite{methods}, yielding $g_0^\text{c}=15.0\pm 0.1$ and $g_0^\text{ab} = 1.5\pm 0.1$ (figure~\ref{figS3}).

At temperatures below ${\sim}1$~K, the magnetic properties of ErTa$_7$O$_{19}$ depart from the single-ion CEF model for $H_0 {\parallel} c$ (Fig.~\ref{fig2}B,C), implying spin--spin interactions of this magnitude.
In contrast, for $H_0 {\perp} c$, the deviations from the CEF model become apparent only below $\sim$0.2~K (Fig.~\ref{fig2}B). 
To estimate the most relevant spin--spin interactions quantitatively, we turn to Weiss molecular-field analysis (Fig.~\ref{fig2}B) of the magnetization at low temperatures~\cite{methods}. 
This reveals dominant antiferromagnetic (AFM) interactions among $c$-axis spin components, corresponding to a Weiss temperature $\theta_\text{CW}^\text{c} = -341\pm 5$~mK, and smaller AFM interactions for $ab$-plane spin components, corresponding to a Weiss temperature $\theta_\text{CW}^\text{ab} = -70\pm 2$~mK.
The derived Weiss temperatures are a combined effect of the anisotropic nearest-neighbor ($nn$) intra-plane exchange interaction $J_\text{e}$ and long-range dipolar interactions $J_{\text{d}i}$ with all spins $i$ (Fig.~\ref{fig1}A). 
These comprise the spin Hamiltonian
\begin{equation}
\mathcal{H}=\mathcal{H}_\text{d} + \sum_{(ij)}\left[J_\text{e}^\text{z} S_i^\text{z}S_j^\text{z} + J_\text{e}^\text{xy} \left(S_i^\text{x}S_j^\text{x} +S_i^\text{y}S_j^\text{y} \right)\right],
\label{exdip}
\end{equation}
containing the dipolar part $\mathcal{H}_\text{d}$, given by Equation~\ref{eq_dip_hamiltonian}, and an XXZ exchange part, where the sum runs over the $nn$ pairs.
The dipolar interactions can be calculated exactly from the crystal structure and $g$ factors~\cite{methods} and are found to be the dominant contribution to the experimental Weiss temperature for $H_0||c$, while they yield only a small correction for $H_0{\perp}c$, since $g_0^\text{c} \gg g_0^\text{ab}$  (table~\ref{tabdipol}).
In particular, the largest interaction is the AFM $nn$ intra-plane dipolar interaction $J_\text{d1}^\text{z}/k_\text{B}=587$~mK, where $k_\text{B}$ is the Boltzmann constant, while both components of the AFM $nn$ intra-plane exchange interaction are much smaller, $J_\text{e}^\text{z}/k_\text{B} = 38\pm 4$~mK and $J_e^\text{xy}/k_\text{B} = 52\pm 3$~mK~\cite{methods}. 
As the dipolar interactions possess strong Ising anisotropy along the $c$ direction, the corresponding spin Hamiltonian given by Equation~\ref{exdip} is quasi-$U(1)$ invariant, which allows for a realization of the supersolid state~\cite{Gao2024double}.
The Ising-like anisotropy on the triangular lattice should for $H {\parallel} c$ lead to a 1/3-magnetization plateau that replaces the spin supersolid state~\cite{yamamoto2014quantum, ulaga2025easy, ulaga2025anisotropic}, which is in line with our experimental observations at 70 mK (Fig.~\ref{fig2}C and figure~\ref{figS6}C).

\subsection*{Magnetic instabilities}
\begin{figure}[h!]
\centering
\includegraphics[trim = 0mm 0mm 0mm 0mm, clip, width=0.95\linewidth]{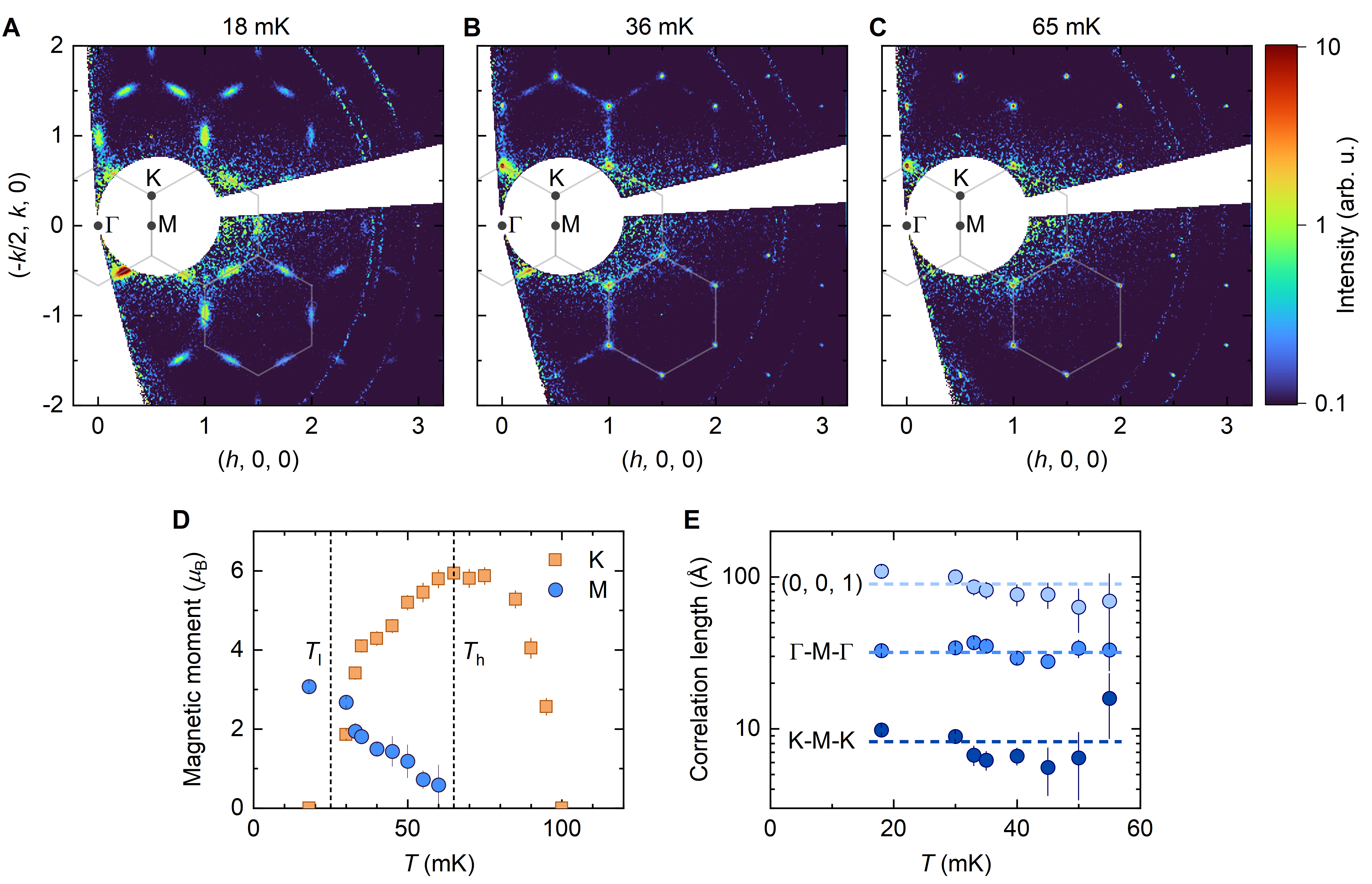}
\caption{{\bf Magnetic correlations in ErTa$_7$O$_{19}$.} 
(\textbf{A--C}) 
Zero-field single-crystal neutron diffraction at various temperatures below the N\'eel temperature $T_\text{N} = 100$~mK showing magnetic Bragg peaks at K points, corresponding to the magnetic propagation vector ${\bf q}_\text{K}=(1/3,1/3,0)$, and a diffuse signal at M points, corresponding to the magnetic propagation vector ${\bf q}_\text{M}=(1/2,0,0)$.
The data taken at 0.8 K were subtracted to remove nuclear Bragg peaks and the background signal from the sample holder
(see figure~\ref{figS3ND} for the raw data). 
Selected Brillouin zone boundaries are highlighted by thin solid lines.
(\textbf{D}) 
The temperature dependence of the long-range ordered (K points) and short-range correlated (M points) magnetic moments, proportional to the square root of the neutron-diffraction intensity\cite{methods}.
The vertical lines highlight the temperatures $T_\text{l}\simeq25$~mK and $T_\text{h}=65$~mK between which the two states coexist.
(\textbf{E}) 
Correlation lengths of the $\text{SR}_\text{M}$ state along three orthogonal directions versus temperature.
The dashed lines are guides to the eye.
}
\label{fig3}
\end{figure}

The magnetization curve exhibits a pronounced anomaly around $T_\text{N} = 100$~mK for $H_0{\parallel} c$ (\ref{fig2}b), indicating a magnetic transition.
This coincides with the establishment of sharp magnetic Bragg peaks appearing in neutron diffraction (ND) experiments below $T_N$ at K points~\cite{methods}---corners of the Brillouin zones (Fig.~\ref{fig3}C and figure~\ref{figS3ND}D)---corresponding to the magnetic propagation vector ${\bf q}_\text{K}=(1/3,1/3,0)$ and thus reflecting a three-sublattice order.
Their width is resolution-limited and yields a lower bound on the correlation lengths that are larger than several hundreds of angstroms along all three directions\cite{methods}.
This clearly demonstrates three-dimensional (3D) long-range ordering, which is also reflected in an anomaly in the specific heat at $T_\text{N}$ (Fig.~\ref{fig2}E).
Refinement of the magnetic structure at 50~mK 
(figure~\ref{figS3ND}F) reveals an out-of-plane Ising-type order $\frac{\rm U}{2}\frac{\rm U}{2}{\rm D}$~\cite{methods}, which breaks translational symmetry and represents the solid component of the spin supersolid.
The amplitude of the superfluid counterpart, corresponding to the ordered in-plane spin component that breaks $U(1)$ symmetry, on the other hand,
cannot be directly determined from our experiments\cite{methods} because of the large $g$-factor anisotropy,$g_0^\text{c}/g_0^\text{ab} \simeq 10$, and the small expected size of this component close to the Ising limit~\cite{wang2009extended, heidarian2010supersolidity, xu2025simulating, ulaga2025easy, gallegos2025phase}.

The long-range ordered state is characterized by enhanced spin fluctuations, as witnessed by the magnetic contribution to the specific heat $c_\text{m}$~\cite{methods}.
The peak in $c_\text{m}$ at $T_\text{N}$ is not very substantial (Fig.~\ref{fig2}E), resulting in only a small change in entropy $S$ across the transition (Fig.~\ref{fig2}F).
This is in sharp contrast to the $c_\text{m}$ data for $H_0{\parallel}c$, where in the field of 0.1~T a much stronger anomaly is observed at a higher $T_\text{N}'=143$~mK (Fig.~\ref{fig2}E), corresponding to significantly larger entropy release (Fig.~\ref{fig2}F) and related to the transition into the ordered 1/3-plateau phase (Fig.~\ref{fig2}C).
Similar enhancements of the transition temperature and entropy release for the field-induced plateau phase compared to the zero-field order were also observed in two well-studied spin supersolid candidates, Na$_2$BaCo(PO$_4$)$_2$~\cite{gao2022spin} and K$_2$Co(SeO$_3$)$_2$~\cite{zhu2024continuum}.

Unexpectedly, on cooling in zero field, the amplitude of the ordered moments develops a maximum of $5.9\pm 0.1\mu_\text{B}$ at $T_\text{h}=65$~mK (Fig.~\ref{fig3}D), which is somewhat reduced from the full-moment value of $g_0^\text{c}\mu_\text{B}/2=7.5\mu_\text{B}$.
The decrease of the ordered moment below $T_\text{h}$ is accompanied by the appearance of diffuse magnetic signal, which progressively grows around the M points (midpoints of the Brillouin zone edges; Fig.~\ref{fig3}B), corresponding to the magnetic propagation vector ${\bf q}_\text{M}=(1/2,0,0)$ and thus to the formation of 
short-ranged stripe correlations.  
The shift of the ND signal is completed at $T_\text{l}\simeq 25$~mK (Fig.~\ref{fig3}D and figure~\ref{figS5}E), where the K-point magnetic Bragg peaks completely disappear (Fig.~\ref{fig3}A) and short-range stripe-correlated magnetic moments prevail.
The magnetic moment at the lowest experimentally accessible temperature of 18~mK amounts to only $3.1\pm0.2\mu_\text{B}$ (see Methods), which is significantly reduced compared to the full-moment value and implies the presence of strong quantum fluctuation in the $\text{SR}_\text{M}$ state.

The ND signal at the M points remains broad at all temperatures. 
The corresponding correlation length within a stripe (K--M--K direction in reciprocal space) is only ${\sim}10$~\AA~($1.5a$), while it is ${\sim}30$~\AA~($5a$) along the stripe propagation direction ($\Gamma$--M--$\Gamma$), and ${\sim}100$~\AA~($5c$) along the $[001]$ out-of-plane direction. 
Furthermore, these correlation lengths are nearly temperature independent (Fig.~\ref{fig3}E), demonstrating that the diffuse nature of the signal at the M points is not due to critical effects usually found when approaching a to long-range ordering transition.
This suggests that the spin--spin correlations should instead remain short-ranged down to zero temperature.
ErTa$_7$O$_{19}$ thus exhibits an unprecedented sequence of magnetic phases, where with decreasing temperature the three-sublattice long-range ordered state $\text{LR}_\text{K}$ is first realized below $T_\text{N}=100$~mK, a coexisting short-range correlated stripe state $\text{SR}_\text{K}$ appears and progressively gets stronger below $T_\text{h}= 65$~mK, and a complete transformation into the latter state is realized below $T_\text{l}\simeq 25$~mK.
This transformation from a long-range correlated state to a short-range correlated state on decreasing temperature is a magnetic analogue of the famous Pomeranchuk effect.

\subsection*{Origin of the spin Pomeranchuk effect}
Trying to understand the peculiar appearance of the magnetic phases in ErTa$_7$O$_{19}$, we first note that for a purely Ising AFM $nn$ interaction $J_\text{e}^\text{z}$ on the triangular lattice, a classical spin-liquid ground state is realized, with an energy per site $E_0/N = - J_\text{e}^\text{z}/4$ and a residual entropy per site $S_0/N = 0.323k_\text{B}$~\cite{wannier1950antiferromagnetism}.
This state, known as the Wannier state, is however unstable against interaction with next-nearest neighbors $J_2$, which induces a stripe-ordered ground state corresponding to the magnetic propagation vector ${\bf q}_\text{M}$~\cite{metcalf1974ground}.
This state has energy per site $E_0/N = -(J_\text{e}^\text{z}+J_2)/4$, 
but at the same time the entropy is removed due to lifting of the macroscopic degeneracy.
However, since the relevant thermodynamic quantity at finite temperatures is the free energy $F = E - T S$, the Wannier state could still be stabilized at elevated temperatures due to its large entropy.

Such a transition has indeed been predicted by Monte Carlo (MC) calculations for the long-range purely dipolar Ising model~\cite{smerald2018spin}, effectively mimicking a mix of $J_\text{e}^\text{z}$ and $J_2$ interactions.
According to these calculations, the ground state is the expected stripe order with long-range correlations at the M points.
On increasing the temperature, a first-order phase transition to a classical spin-liquid state with finite entropy occurs at 
$T_\text{1} = 0.046 J_\text{1}/k_\text{B}$, where $J_\text{1}$ is the $nn$ coupling.
Above $T_\text{1}$, the spin structure factor is first distributed around the Brillouin zone boundary, but the weight of the correlations progressively shifts to the K points around $T_2=0.086J_\text{1}/k_\text{B}$.
The gradual shift of magnetic correlations in ErTa$_7$O$_{19}$ from M to K points on increasing the temperature between $T_\text{l}$ and $T_\text{h}$, thus resembles the predictions of the dipolar Ising model.
This is not surprising, given the dominance of the Ising-type dipolar interactions (table~\ref{tabdipol}).
However, while the classical model predicts a long-range ordered stripe ground state with correlations at the M points and a short-range ordered spin-liquid state with correlations at the K points at higher temperatures, our experiments on the contrary show a short-range correlated stripe ground state, $\text{SR}_\text{M}$, and a long-range ordered three-sublattice state, $\text{LR}_\text{K}$, at elevated temperatures.

We attribute this striking difference between the classical model and our experiments to the presence of additional interactions that introduce quantum effects.
The transverse component of the $nn$ exchange interaction $J_\text{e}^\text{xy}$ (Equation~\ref{exdip}) induces quantum fluctuations that favor the formation of local singlet spin pairs~\cite{ulaga2024finite}.
As a result, the transition temperature to the spin-stripe ground state, $T_\text{1}$, should be reduced compared to the classical model.
In fact, a large enough $J_\text{e}^\text{xy}$ completely suppresses the stripe phase and stabilizes the long-range ordered spin supersolid phase~\cite{wang2009extended, heidarian2010supersolidity, yamamoto2014quantum, sellmann2015phase, gao2022spin, Gao2024double, xu2025simulating, ulaga2025easy} with the magnetic propagation vector ${\bf q}_\text{K}$ instead~\cite{ulaga2024finite}.

The quantum destabilization of the stripe phase is indeed confirmed by fully quantum finite-temperature Lanczos (FTLM) method calculations of 
a 2D model that incorporates both long-range Ising and $nn$ $J_\text{e}^\text{xy}$ interactions~\cite{methods}.
The long-range interactions decay with distance $r$ (in lattice-spacing units) as $J_1/r^\kappa$, where $\kappa = 3$ corresponds to a purely dipolar model without the $nn$ Ising exchange component $J_\text{e}^\text{z}$, while $\kappa > 3$ mimics a finite AFM $J_e^z$.
In this model, the total $nn$ Ising interaction is related to the parameters in Equation~\ref{exdip} via $J_1 = J_\text{e}^\text{z} + J_{\text{d}1}^{z}$, and the anisotropy parameter is $\alpha=J_\text{e}^\text{xy}/J_1$ ($\alpha=0.08$ for ErTa$_7$O$_{19}$; table~\ref{tabdipol}).

\begin{figure}[h!]
\centering
\includegraphics[trim = 0mm 0mm 0mm 0mm, clip, width=1\linewidth]{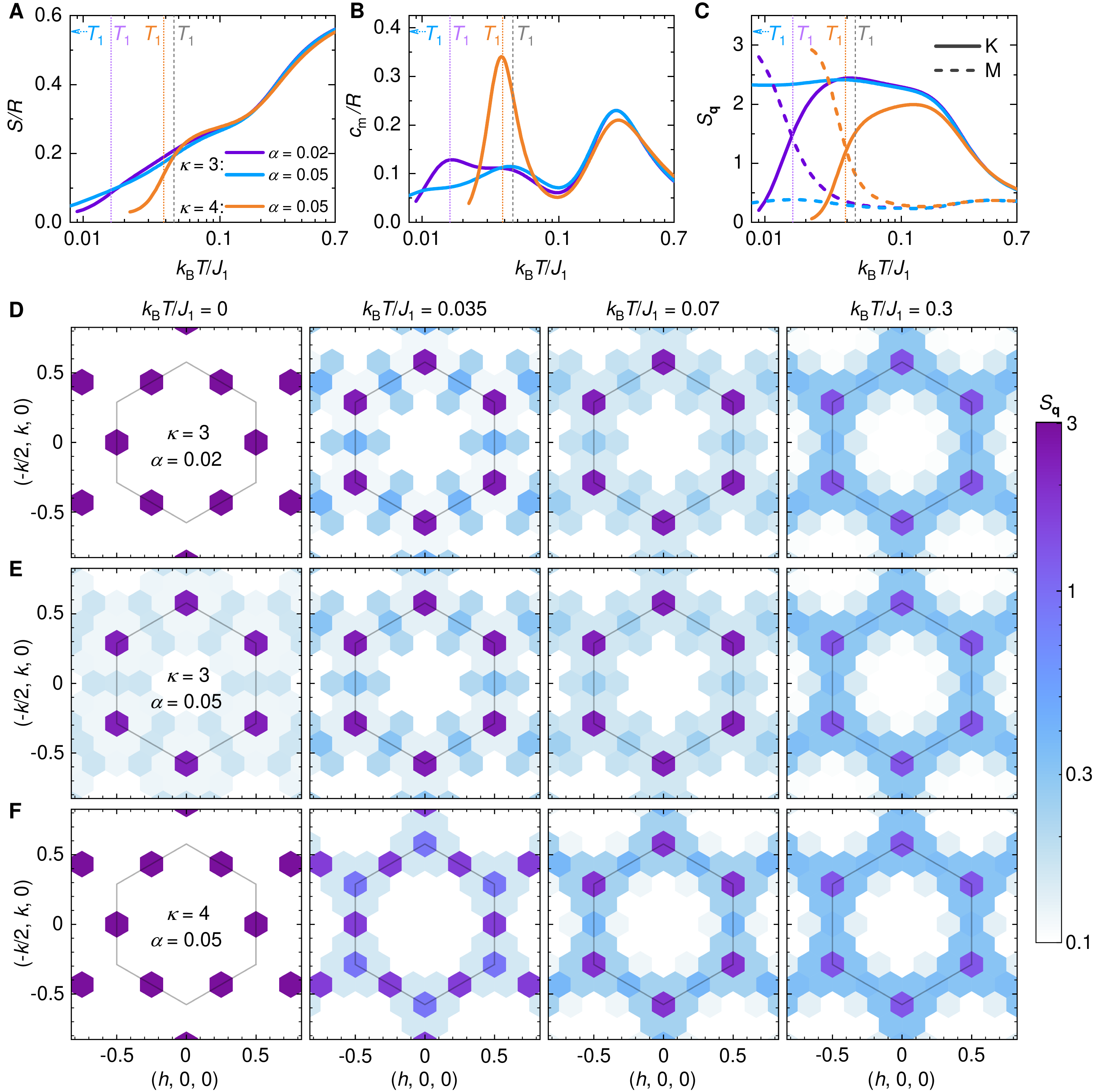}
\caption{{\bf Thermodynamic properties and spin structure factor.} 
(\textbf{A,B})  
The temperature dependence of the magnetic entropy $S$ and magnetic specific heat $c_\text{m}$ calculated~\cite{methods} using the finite-temperature Lanczos method (FTLM) at different transverse anisotropies $\alpha = J_\text{e}^\text{xy}/J_{1}$ for pure dipolar interactions ($\kappa = 3$) and modified long-range interactions ($\kappa = 4$).
(\textbf{C})  
The temperature dependence of the static spin structure factor $S_{\bf q}$ at ${\bf q}_\text{K}=(1/3,1/3,0)$ and ${\bf q}_\text{M}=(1/2,0,0)$ calculated via FTLM.
The colored dotted vertical lines indicate the transition temperature $T_\text{1}$ between the M and K states obtained from FTLM for different $\kappa$ and $\alpha$, while the arrow indicates 
$T_\text{1} \rightarrow 0$.
The gray dashed vertical line corresponds to the $T_\text{1}$ of the classical ($\alpha = 0$) purely dipolar ($\kappa = 3$) Ising model predicted by Monte Carlo simulations~\cite{smerald2018spin}.
(\textbf{D--F})  
Color plots of the static spin structure factor $S_{\bf q}(T)$ in reciprocal space at a few selected temperatures for different $\kappa$ and $\alpha$.
}
\label{fig4}
\end{figure}

In the purely dipolar model ($\kappa = 3$) the calculations reveal that thermodynamic quantities, such as magnetic entropy $S$ and specific heat $c_\text{m}$, as well as static spin structure factor $S_{\bf q}$, where ${\bf q}$ is a wave vector, heavily depend on $\alpha$ at low temperatures, even for very small anisotropies (Fig.~\ref{fig4}).
For $\alpha = 0.02$, a transition is observed at 
$T_\text{1} \simeq 0.016 J_\text{1}/k_B$, as evidenced in Fig.~\ref{fig4}B by the low-temperature maximum in $c_\text{m}(T)$, which is much lower than the transition temperature $T_\text{1} = 0.046 J_1/k_\text{B}$ between the M and K phases predicted by classical ($\alpha = 0$) dipolar Ising MC calculations~\cite{smerald2018spin}.
The nature of this transition, however, remains the same, as witnessed by a weight shift in the spin structure factor at ${\sim}T_1$ (Fig.~\ref{fig4}C) in the quantum case.
Notably, at $T \gtrsim T_\text{1}$, significant correlations coexist at both the K and the M points (Fig.~\ref{fig4}D). 
Increasing the anisotropy to a still moderate $\alpha = 0.05$ 
dramatically affects the spin correlations, as now $S_{{\bf q}_\text{K}}$ remains dominant over $S_{{\bf q}_\text{M}}$ over the whole temperature range (Fig.~\ref{fig4}C) and the low-temperature peak in $c_\text{m}(T)$ gets suppressed (Fig.~\ref{fig4}B). 
This reveals the stabilization of the three-sublattice spin solid down to the lowest temperatures ($T_\text{1} \rightarrow 0$). 
Interestingly, however, the correlations are present also at the M points, even at zero temperature (Fig.~\ref{fig4}E).  
Going beyond the purely dipolar model, we see that an effective additional $nn$ Ising exchange coupling $J_\text{e}^\text{z}$, introduced with $\kappa > 3$, acts against the anisotropy, as it re-establishes the stripe state at low temperatures (Fig.~\ref{fig4}F).
For $\alpha = 0.05$ and $\kappa = 4$ the transition temperature rises to
$T_\text{1} = 0.039 J_1/k_\text{B}$ (Fig.~\ref{fig4}C), which is quite close to the prediction of the classical purely dipolar Ising MC calculations~\cite{smerald2018spin}.

At intermediate temperatures $T_1 < T \lesssim J_1/k_B$, our modeling predicts correlations building up across the Brillouin zone boundary, which are peaked at the K points (Fig.~\ref{fig4}D--F).
The three-sublattice long-range ordering observed experimentally in ErTa$_7$O$_{19}$ below $T_\text{N}=100$~mK is a consequence of an effective ferromagnetic (FM) interlayer interaction of dipolar origin, $J^\text{MF}_{{\bf q}_\text{K}}/J_{1} = -0.20$, determined via a mean-field approach~\cite{methods}.
The estimated value of the N\'eel transition temperature, $T_\text{N}\sim 0.3 J_{1}/k_\text{B} \sim 0.2$~K, is somewhat overestimated, as expected, since the mean-field approach neglects spin fluctuations.
The same approach also predicts long-range ordering into the stripe phase around $T_1$, below which $S_{{\bf q}_\text{M}}>S_{{\bf q}_\text{K}}$ (Fig.~\ref{fig4}C).

The fact that ErTa$_{7}$O$_{19}$ enters a short-range correlated state at the lowest temperatures after exhibiting the symmetry-breaking long-range ordering at higher temperatures defies the mean-field prediction.
The short-range nature of correlations is most likely due to enhanced quantum fluctuations in the ground state. 
Disordered QSL states that result from such fluctuations are regularly encountered in extended regions of phase diagrams of triangular-lattice antiferromagnets on the borders between spin supersolid and stripe phases; for instance, in the case of the XXZ model with further-neighbor interactions~\cite{gallegos2025phase}, or when additional bond-dependent off-diagonal exchange anisotropy terms~\cite{zhu2018topography} are present that lead to Kitaev-type interactions~\cite{ortiz2023quantum}.
The unconventional transitions in ErTa$_7$O$_{19}$ from the long-range ordered three-sublattice spin solid to the short-range correlated spin-stripe phase upon cooling can thus be attributed to a complex interplay of long-range dipolar interactions and anisotropic $nn$ exchange interactions, inducing quantum fluctuations.

\subsection*{Outlook}
The spin Pomeranchuk effect in quantum magnets reveals a conceptually rich interplay of energetic, entropic, and quantum effects, and is expected to be universal in frustrated spin systems.  
As the effect is tunable via temperature and applied magnetic field, it offers a powerful route to stabilizing novel exotic magnetic phases by entropy engineering and inducing transformations among them.
Its experimental realization in ErTa$_7$O$_{19}$, for example, demonstrates the emergence of a complex long-range ordered state at higher temperatures with unusually high entropy, despite the short-range correlated ground state of the material. 
Beyond opening up new avenues in fundamental research, the spin Pomeranchuk effect also provides new opportunities for technological applications.
The entropy-stabilized phases could be exploited for efficient adiabatic-demagnetization cooling schemes based on enhanced magnetocaloric responses, arising from large entropy differences between competing magnetic phases.
This may prove particularly valuable in cases where conventional cryogenics is inadequate or challenging. 
For example, in quantum devices that require efficient local cooling, such magnets could be used as on-chip field-driven refrigerants.
Moreover, the large changes in the magnetic state in the spin Pomeranchuk transition, enabled by relatively small energy inputs, point toward energy-efficient magnetic functionalities that are promising for spintronics, memory devices, and related applications.



\clearpage 

%
\bibliography{ETO_paper_SCI} 
\bibliographystyle{sciencemag}

%
%
%
%
%
%


\section*{Acknowledgments}
We thank Fr\'ed\'eric Mila and Wei Li for enlightening discussions.
\paragraph*{Funding:}
We acknowledge the financial support of the Slovenian Research and Innovation Agency through the Programs No.~P1-0125 and P2-0348, and Projects No.~N1-0148, J1-50008, BI-US/22-24-065, N1-0356, J1-50012, and J2-60034.
P.K. acknowledges the funding by the Anusandhan National Research Foundation (ANRF), Department of Science and Technology, India through Research Grants.
N.B acknowledges the Program QuanTEdu-France n° ANR-22-CMAS-0001 France 2030.
INS and ND experiments at the ISIS Neutron and Muon Source were supported by beam-time allocations RB2220319 and RB2420338, respectively, both approved by the Science and Technology Facility Council.
A portion of this work was performed at the National High Magnetic Field Laboratory, which is supported by National Science Foundation Cooperative Agreement No.~DMR-2128556 and the State of Florida.
\paragraph*{Author contributions:}
A.Z.~is the corresponding author who conceived the project, designed and supervised the experiments, as well as wrote the paper with feedback from all co-authors. 
P.K.~is the corresponding author who initiated magnetic studies of rare-earth heptatantalates, and ErTa$_7$O$_{19}$ in particular. 
P.K., U.J., and B.S.~synthesized polycrystalline samples and performed the Rietveld refinement of powder XRD data.
P.K.~and U.J.~also conducted the specific heat measurements.
Single-crystal samples were grown by L.\v{S}.~under the supervision of M.D., who both structurally characterized the crystals.
Z.J., N.B, and E.L.~performed the bulk magnetic measurements.
A.Z.~and J.v.T.~conducted the ESR study.
K.J., M.P., T.A., P.M., F.O., and D.K.~conducted and analyzed the neutron diffraction measurements.
M.D.L.~and T.A.~conducted the INS measurements, the corresponding CEF modeling was performed by K.\v{Z}.~and M.P.,
who also performed reverse MC refinements of diffuse magnetic scattering.
M.M.~and M.G.~performed DFT calculations of the EFG tensor that were used to estimate the nuclear specific heat.
M.G.~also performed dipolar-interaction calculations.
M.U.~and P.P.~conducted the FTLM calculations, P.P.~also carried out the MF modeling of 3D ordering.
\paragraph*{Competing interests:}
There are no competing interests to declare.
\paragraph*{Data and materials availability:}
All data are available in the manuscript or the supplementary materials. The raw inelastic neutron scattering data can be found in Ref.~\cite{INS} and the raw neutron diffraction data in Ref.~\cite{ND}.


\subsection*{Supplementary materials}
Materials and Methods\\
Figs. S1 to S6\\
Tables S1 to S4\\
References \textit{(41-\arabic{enumiv})}\\ 


\newpage


\renewcommand{\thefigure}{S\arabic{figure}}
\renewcommand{\thetable}{S\arabic{table}}
\renewcommand{\theequation}{S\arabic{equation}}
\renewcommand{\thepage}{S\arabic{page}}
\setcounter{figure}{0}
\setcounter{table}{0}
\setcounter{equation}{0}
\setcounter{page}{1} 


\begin{center}
\section*{Supplementary Materials for\\ \scititle}

   K.~Jakseti\v c$^{1,2}$,
	T.~Arh$^{1,2,3}$,
	M.~Pregelj$^{1,2}$,
    M.~Gomil\v{s}ek$^{1,2}$,
    M.~Dragomir$^{1}$,\\
    P.~Prelov\v{s}ek$^{1}$, 
    M.~Ulaga$^{4}$,
    L.~\v{S}ibav$^{1}$, 
    M.~Malovrh$^{2}$, 
    K.~\v{Z}eleznikar$^{2}$, 
    Z.~Jagli\v{c}i\'c$^{5,6}$,\\
    P.~Manuel$^{7}$,
    F.~Orlandi$^{7}$, 
    D.~Khalyavin$^{7}$, 
    M.~D. Le$^{7}$,
    N.~Bujault$^{8}$, 
    E.~Lhotel$^{8}$,\\
    J.~van~Tol$^{9}$, 
    U.~Jena$^{10}$, 
    B.~Sana$^{10}$, 
    P.~Khuntia$^{10,11\ast}$, 
    and A.~Zorko$^{1,2\ast\ast}$\\ 
\small$^\ast$Corresponding author. Email: pkhuntia@iitm.ac.in \\
    \small$^{\ast\ast}$Corresponding author. Email: andrej.zorko@ijs.si
\end{center}

\subsubsection*{This PDF file includes:}
Materials and Methods\\
Figures S1 to S6\\
Tables S1 to S4\\

\newpage


\subsection*{Materials and Methods}

\begin{figure}[h!]
\centering
\includegraphics[width=1\linewidth]{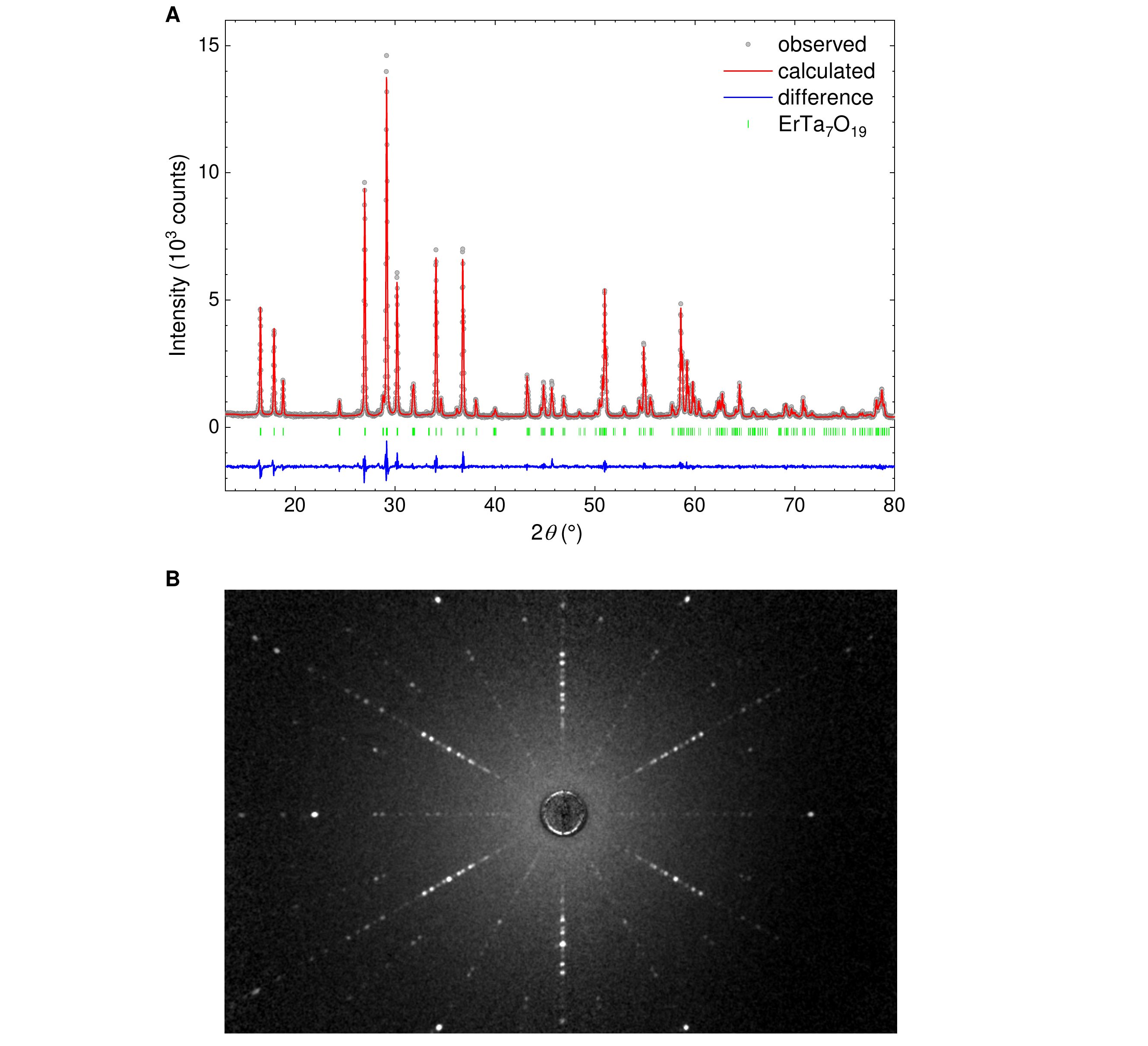}
\caption{{\bf Structural refinement of ErTa$_7$O$_{19}$.} 
(\textbf{A})  
Rietveld refinement of powder XRD data taken at room temperature within the space group P$\bar{6}c2$  ($R_F=2.5$).
(\textbf{B})  
Laue diffraction pattern of a typical ErTa$_7$O$_{19}$ single crystal.
}
\label{figS1}
\end{figure}

{\bf\noindent Synthesis and Crystal structure}\\
A solid-state method was employed to synthesize high-quality polycrystalline ErTa$_7$O$_{19}$ and a flux method was optimized to grow single crystals of sizes up to $1.5\times1.5\times0.5$~mm$^3$, as detailed in Ref.~\cite{sibav2025optimized}.
Powder X-ray diffraction (XRD) was performed at room temperature using a Rigaku X-ray diffractometer working with the Cu $K_\alpha$ radiation. 
Structural refinement (Fig.~\ref{figS1}A) was performed within the space group No.~188 (P$\bar{6}$c2), as previously done for NdTa$_7$O$_{19}$~\cite{arh2022ising}.
The corresponding atomic positions and lattice parameters are summarized in Table~\ref{struct}.
The quality of single crystals used in our study was verified by Laue diffraction, with a typical example shown in Fig.~\ref{figS1}C.
Their crystal structure was further checked via ND performed on a 20-mg single crystal using WISH time-of-flight diffractometer at ISIS, UK (Fig.~\ref{figS3ND}E).
The structure was found to remain unchanged with temperature (Table~\ref{struct}).

\begin{table}[t]
\centering
\caption{
{\bf Structural parameters of ErTa$_7$O$_{19}$.}
Atomic positions obtained from Rietveld refinement of powder XRD data at room temperature and of single-crystal neutron diffraction (ND) data at 50~mK (Space Group No. 188, $P\bar{6}c2$) with $a=b=6.20178(3)$~\r{A}, $c=19.85693(5)$~\r{A}, $\alpha=\beta=90^\circ$, and $\gamma=120^\circ$.\\
}
\begin{tabular}{c|c|c c c c c} 
 \hline
 Atom & Wyckoff & $x$ & $y$ & $z$ & ${\rm B_{iso}}$ & Occ. \\
 \hline
 & & \multicolumn{5}{c}{XRD}\\
 Er & 2$c$ & 1/3 & 2/3 & 0 & 0.2(1) & 1  \\
 Ta1 & 2$e$ &2/3& 1/3& 0& 0.4(1) & 1 \\
 Ta2& 12$l$ &0.307(6)2& 0.3586(9)& 0.1553(1)& 0.5 & 1 \\
 O1& 12$l$ &0.224(3)& 0.967(5)& 0.1569(6)& 0.5 & 1 \\
 O2& 12$l$ &0.406(3)& 0.036(5)& 0.9465(5)&1.2(4) & 1 \\
 O3& 6$k$ &0.415(3)& 0.367(5)& 1/4& 0.6(5)& 1 \\
 O4& 4$i$ &2/3& 1/3& 0.1673(1)& 0.5& 1 \\
 O5& 4$h$ &1/3& 2/3& 0.128(1)& 0.6(8)& 1 \\
 \hline
 & & \multicolumn{5}{c}{ND}\\
 Er & 2$c$ & 1/3 & 2/3 & 0 &/ & 1 \\
 Ta1 & 2$e$ &2/3& 1/3& 0&/ & 1\\
 Ta2& 12$l$ &0.363(2)& 0.364(2)& 0.1554(2)&/ & 1\\
 O1& 12$l$ &0.246(2)& 0.998(5)& 0.1527(2)&/ & 1\\
 O2& 12$l$ &0.427(1)& 0.044(1)& 0.9445(2)&/ & 1\\
 O3& 6$k$ &0.374(2)& 0.423(2)& 1/4&/ & 1\\
 O4& 4$i$ &2/3& 1/3& 0.1654(4)&/ & 1\\
 O5& 4$h$ &1/3& 2/3& 0.1274(5)&/ & 1\\
 \hline
\end{tabular}
\label{struct}
\end{table}

\subsubsection*{Bulk magnetic characterization}
Bulk magnetic characterization of a 13-mg single crystal was done using a Quantum Design SQUID magnetometer in the temperature range between 2 and 300~K and a custom-built SQUID magnetometer in the temperature range between 70 mK and 4.2 K at the Institut N\'eel, Grenoble.
In the latter experiment, the crystal was glued on a copper sample holder with Apiezon N grease to ensure good thermal contact.
The temperature dependence of the magnetization ($M$) was measured in applied fields of $\mu_0 H_0 = 5$, 10 and 100~mT, while isothermal magnetization curves were measured at selected temperatures in fields up to 8~T.
In Fig.~\ref{fig2}A the magnetization data are normalized by the internal field $H_\text{i}=H_0 - NM$, given by the demagnetization factors $N_\text{c} =0.57$ and $N_\text{ab} =0.17$. 
These factors were determined from the dimensions of the plate-like sample, which 
was approximated by a parallelepiped of size $2.26\times 1.52\times 0.66$~mm$^3$.
At the lowest temperatures, the correction due to demagnetization was found to be as large as $N M/H_0=0.61$ for fields along the $c$ direction and 0.017 for fields perpendicular to $c$.
No field-cooled--zero-field-cooled (FC--ZFC) splitting was observed in magnetic measurements.

\begin{figure}[h!]
\centering
\includegraphics[width=1\linewidth]{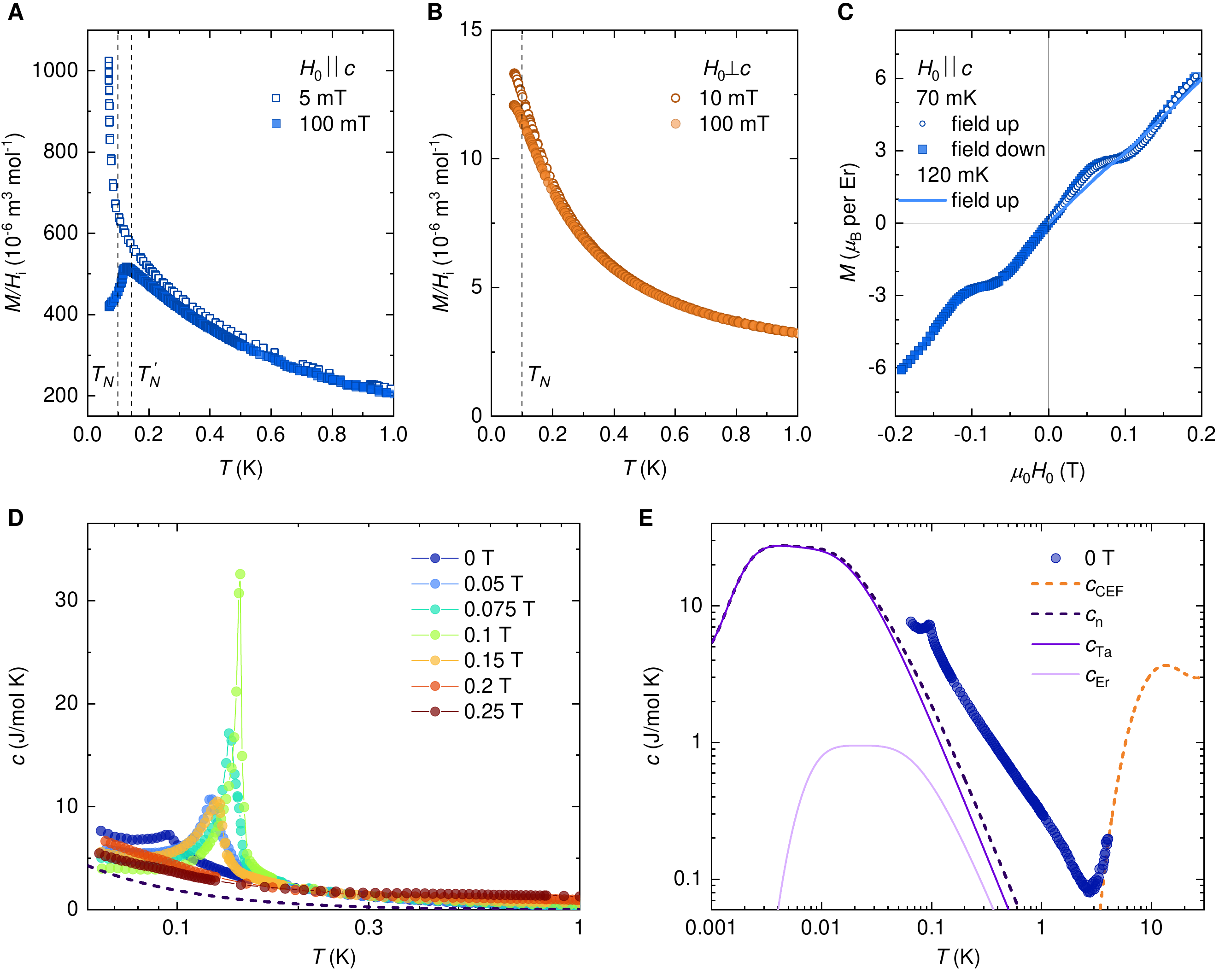}
\caption{{\bf Low-temperature bulk properties of ErTa$_7$O$_{19}$.} 
(\textbf{A,B}) 
The low-temperature normalized magnetization $M/H_\text{i}$ for the external field $H_0$ applied in the direction perpendicular to the triangular planes ($H_0 {\parallel}c$) and within the planes ($H_0 {\perp}c$). 
$M$ denotes magnetization and $H_\text{i}=H_0 - NM$ the internal field, given by the demagnetization factors $N_\text{c} =0.57$ and $N_\text{ab} =0.17$.
The vertical lines show the N\'eel transition temperature $T_\text{N} = 100$~mK in zero field and the 1/3-plateau ordering $T_\text{N}' = 143$~mK in the field $\mu_0 H_0 =100$~mT.
(\textbf{C}) 
Isothermal magnetization curve at 70 and 120~mK.
The 70-mK data were recorded for increasing the field from zero to $\mu_0H_{\rm max}=0.1925$~T (field up) and for decreasing the field from  $\mu_0H_{\rm max}$ to $-\mu_0H_{\rm max}$ (field down).
(\textbf{D}) 
The specific heat in zero and selected fields $H_0 {\parallel}c$.
The dashed line shows the nuclear contribution $c_\text{n}$ obtained from DFT calculations~\cite{methods}.
(\textbf{E}) 
The experimental specific heat measured in zero field compared to the contribution due to excited Kramers doublets $c_\text{CEF}$ and the nuclear contribution $c_\text{n}$.
The latter is a combined contribution from $^{181}$Ta and $^{167}$Er nuclei, while that of $^{17}$O nuclei is negligible.
}
\label{figS6}
\end{figure}

The $M/H_\text{i}$ data at different fields are shown in Fig.~\ref{figS6}.
At temperatures slightly above $T_\text{N}=100$~mK, the magnetization is linear in applied fields up to at least 0.2~T (Fig.~\ref{figS6}C). 
We, therefore, analyzed the data within a Weiss molecular-field approximation appropriate for interacting spins in the Kramers ground state doublet~\cite{pocs2021systematic},
\begin{equation}
\frac{1}{\chi}=\frac{1}{C}\left(\frac{T}{1 -a T} - \theta_\text{CW} \right),
\label{eqCW}
\end{equation}
where $ C=N_\text{A}\mu_0\mu_\text{B}^2g^2/(4k_\text{B})$ is the Curie constant ($N_\text{A}$ is the Avogadro number, $\mu_0$ the vacuum permeability, $\mu_\text{B}$ the Bohr magneton, $g$ the $g$ factor, and $k_\text{B}$ the Boltzmann constant), the $a T$ term is the Van Vleck correction accounting for the second-order perturbative effect of the Zeeman Hamiltonian, and the Weiss temperature $\theta_\text{CW}$ is an effective measure of the spin--spin interactions within the Kramers doublet.
The fitting of the model to the data (Fig.~\ref{fig2}B) in the temperature range between 0.15 and 4~K yielded $g$ factors $g_0^\text{c}=15.1\pm 0.1$ and $g_0^\text{ab} = 1.4\pm 0.1$, and Weiss temperatures $\theta_\text{CW}^\text{c} = -341\pm 5$~mK and $\theta_\text{CW}^\text{ab} = -70\pm 2$~mK for the field directions parallel and perpendicular to the $c$ axis, respectively.
The Van Vleck terms were found to be $a_\text{c} = 0$ and $a_\text{ab} = -0.56\pm 0.01$~K$^{-1}$. 
Fitting in this temperature range assured that critical effects arising close to $T_\text{N} = 100$~mK (see below) were excluded and that the first excited Kramers doublet (at $\Delta = k_\textbf{B}\cdot32.5$~K, see below) did not noticeably affect the magnetization. 
The latter was tested by adding additional terms originating from excited Kramers doublets~\cite{pocs2021systematic}, which did not affect the ground-state parameters.
For both directions, the experimental data deviates from the model at $T\lesssim |\theta_\text{CW}|$, implying a limitation of the mean-field approach at the lowest temperatures.

\subsubsection*{CEF splitting}
Inelastic neutron scattering (INS) was carried out to detect the splitting of the Er$^{3+}$ ground-state multiplet $^{4}I_{15/2}$ in  ErTa$_7$O$_{19}$ into eight Kramers doublets due to crystal electric field (CEF).
The experiment was conducted at 5~K on a 6.4-g polycrystalline sample using MARI time-of-flight neutron spectrometer at the ISIS Pulsed Neutron and Muon Source of the Rutherford Appleton Laboratory, UK~\cite{INS}.
The sample was put in a hollow cylindrical aluminum sample can with an outer diameter of 5~cm.
Measurements were performed with neutron incident energies of 29.7, 70, and 180 meV, using a Fermi chopper system with a Gd foil chopper pack rotating at 400~Hz.

\begin{figure}[h!]
\centering
\includegraphics[trim = 0mm 0mm 0mm 0mm, clip, width=\linewidth]{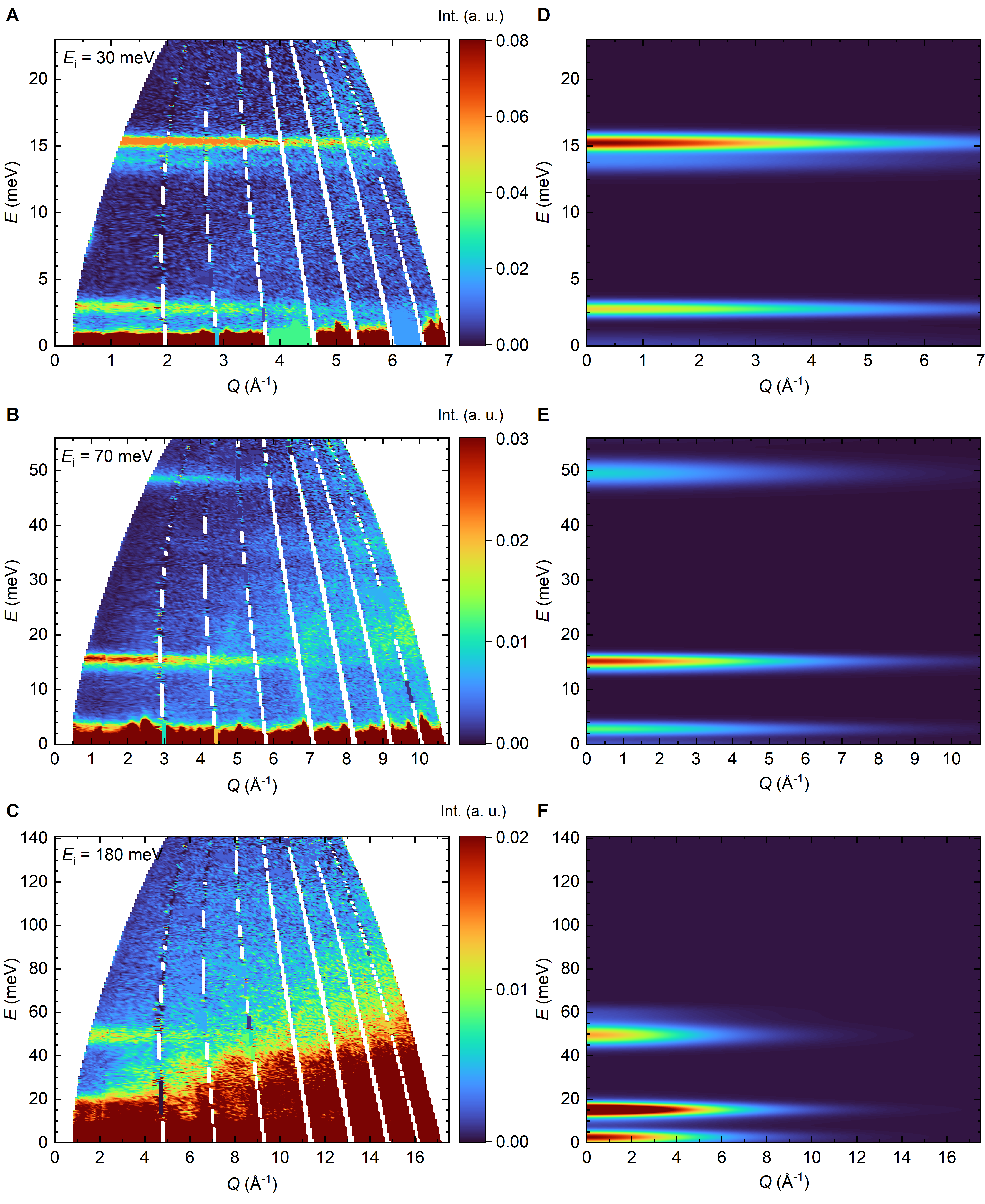}
\caption{{\bf Crystal electric field and phonon excitations in ErTa$_7$O$_{19}$.} 
(\textbf{A--C})  
The spin structure factor measured at 5~K at three different neutron incident energies $E_\text{i}$.
(\textbf{D--F})  
The corresponding predictions based on the CEF model (Equation~\ref{eqCEF}).
}
\label{figS2}
\end{figure}

Several flat INS modes were found experimentally; at 2.8, 13.9, 15, and around 50~meV, while no additional flat levels were observed at higher energies (Fig.~\ref{figS2}A--C).
These were recognized as dispersionless CEF modes, the intensity of which characteristically decays with $Q$, contrary to the dispersed phonon modes, the intensity of which increases with $Q$.
By using \textsc{Phi} software~\cite{chilton2013phi}, we modeled the position and intensity of these modes (Fig.~\ref{figS2}D--F) simultaneously with the temperature-dependent magnetization (Fig.~\ref{fig2}A) and isothermal magnetization curves (Fig.~\ref{fig2}C,D) at temperatures above 2~K to avoid the effect of spin--spin interactions (see below) using the CEF Hamiltonian

\begin{equation}
    \mathcal{H}_{\mathrm{CEF}} = \sum_{l,m} \sigma^l B_l^m \Theta_l \hat{O}_l^m,
    \label{eqCEF}
\end{equation}
where $\sigma^l$ is the orbital reduction parameter to the power of $l$, $B_l^m$ are scaling parameters, $\Theta_l$ are the operator equivalent factors~\cite{chilton2013phi}, and $\hat{O}_l^m$ correspond to Stevens operators~\cite{stevens1952matrix}.
The model works very well in describing all these experiments. 
The scaling parameters allowed by the dihedral $D_3$ point symmetry at the Er$^{3+}$ site, deduced from the fitting, are
$B_2^0 = 9.718$~meV, $B_4^0 =-1.458$~meV, $B_4^3 =-375.7$~meV, $B_6^0 =7.023$~meV, $B_6^3 =24.22$~meV, $B_6^6 =19.97$~meV, and the orbital reduction parameter is $\sigma=1.0492$.
The modelling revealed single excited Kramers doublets at 2.8, 13.9, and 15.0~meV, and four additional doublets at 50--60~meV. 
In this last group, one Kramers doublet is far more intense than the other three (Fig.~\ref{figS2}D).
The composition of all Kramers doublets and their corresponding $g$ factors are summarized in Table~\ref{CEF-vec}.
The extracted $g$ factors of the ground-state and first excited Kramers doublets are $g_0^\text{c}=14.9\pm 0.2$, $g_0^\text{ab}=1.2\pm 0.2$ and $g_1^\text{c}=2.5\pm 0.1$, $g_1^\text{ab}=8.8\pm 0.2$, respectively.
Very similar values were also deduced directly from electron spin resonance spectra (see below).

\begin{table}[t]
\centering
\caption{
{\bf Kramers doublets in ErTa$_7$O$_{19}$.}
The eigenstates $\pm\omega_k$ ($k$\,=\,0-7) of the CEF Hamiltonian given by Equation~(\ref{eqCEF}) in the $|\pm m_J\rangle$ basis, the corresponding energies, and the principal $g$ factors of the $^{4}I_{15/2}$ Er$^{3+}$ multiplet in ErTa$_7$O$_{19}$.\\
\label{CEF-vec}}
\begin{tabular}{c|c c c c c c c c}
 \hline
 $|\pm m_J\rangle$  & $\pm\omega_{0}$ & $\pm\omega_{1}$ & $\pm\omega_{2}$ & $\pm\omega_{3}$ & $\pm\omega_{4}$ & $\pm\omega_{5}$ & $\pm\omega_{6}$ & $\pm\omega_{7}$ \\
\hline
$|\pm15/2\rangle$ &     	0	  &     	0	  & $\pm    	0.015	  $ & 	0	 &     	0.927	  &     	0	  & $\pm	0.374	$ &     	0	\\
$|\pm13/2\rangle$ &     	0.956	  & -0.247	 &     	0	  & $\mp 0.000	$  &	0	&     	0.007	  &     	0	  &    -0.126	\\
$|\pm11/2\rangle$ & $\mp	0.001	$ & $\mp    	0.013	 $ &     	0	  & -0.891	  & 	0	  & $\pm	0.434	$ &     	0	  & $\mp	0.026	$ \\
$|\pm9/2\rangle$  &     	0	  &     	0	  &	0.287	  &     	0	  &  $\mp   	0.339	  $ &     	0	  &     	0.828	  &     	0	\\
$|\pm7/2\rangle$  & $\mp	0.066	$ &  $\pm   	0.289	  $ &     	0	  &     	0	  &     	0	  & $\pm	0.032	$ &     	0	  & $\mp	0.871	$ \\
$|\pm5/2\rangle$  &     	0.001	  & -0.037	 &     	0	  & $\pm    	0.447	 $ &	0	 &     	0.817	  &     	0	  &    -0.096	\\
$|\pm3/2\rangle$  &     	0	  &     	0	  &    $\mp	0.782	$ & 	0	 &  -0.148	  &     	0	  & $\pm	0.398	$ &     	0	\\
$|\pm1/2\rangle$  &    -0.280	  & -0.882	&     	0	  & $\pm    	0.000	$ & 	0	 &     	0.040	  &     	0	  &    -0.187	\\
$|\mp1/2\rangle$  & $\pm	0.010	$ &  $\pm   	0.169	  $ &     	0	  &     	0.034	  &     	0	  & $\pm	0.270	$ &     	0	  & $\pm	0.068	$ \\
$|\mp3/2\rangle$  &     	0	  &     	0	  & -0.441	 &    	0	  & $\pm    	0.048	$ &     	0	  &     	0.101	  &     	0	\\
$|\mp5/2\rangle$  & $\pm	0.031	$ &  $\mp	0.193	 $ &     	0	  & -0.002	  &    	0	  & $\mp	0.120	$ &     	0	  & $\mp	0.262	$ \\
$|\mp7/2\rangle$  &    -0.002	  &	0.055	 &     	0	  & $\mp    	0.051	$ & 	0	  &    -0.215	  &     	0	  &    -0.318	\\
$|\mp9/2\rangle$  &     	0	  &     	0	  & $\pm    	0.332	$ & 	0	 &     	0.035	  &     	0	  & $\pm	0.070	$ &     	0	\\
$|\mp11/2\rangle$ &     	0.037	  &	0.070	 &     	0	  & $\mp    	0.003	$  & 	0	 &     	0.064	  &     	0	  &     	0.070	\\
$|\mp13/2\rangle$ & $\mp	0.033	$ &    $\pm	0.047	 $ &     	0	  &     	0.033	  &     	0	  & $\pm	0.051	$ &     	0	  & $\pm	0.046	$ \\
$|\mp15/2\rangle$ &     	0	  &     	0	  & -0.000	 &   	0	  & $\mp    	0.001	  $ &     	0	  &    -0.001	  &     	0	\\
\hline
$E$(meV) & 0 & 2.8 & 13.9 & 15.2 & 49.5 & 49.7 & 56.6 & 60.0          \\
\hline
$g^\text{c}$ & 14.9 & 2.5 & 4.9 & 12.0 & 17.3 & 6.3 & 10.8 & 7.2          \\
$g^\text{ab}$ & 1.2 & 8.8 & 0 & 0.03 & 0 & 4.3 & 0 & 4.5  \\
 \hline
\end{tabular}
\end{table}

\subsubsection*{ESR spectroscopy}
Electron spin resonance (ESR) spectra were recorded in X-band (at 9.3~GHz) on a commercial Bruker E500 spectrometer and at 120~GHz on a custom-built spectrometer with heterodyne detection at the National High Magnetic Field Laboratory, Tallahassee, Florida.
In both cases, He-flow cryostats were used for cooling the sample.
In X-band, a 1-T electromagnet was used, while the additional field modulation of 0.5~mT was applied at 100~kHz.
At high frequencies, a 12-T sweepable superconducting magnet was used.
Here, the recorded spectra were distorted due to strong microwave absorption at low temperatures, even for the smallest available samples and lowest microwave power; therefore, we only show spectra at an intermediate temperature of 70~K.

\begin{figure}[h!]
\includegraphics[width=1\linewidth]{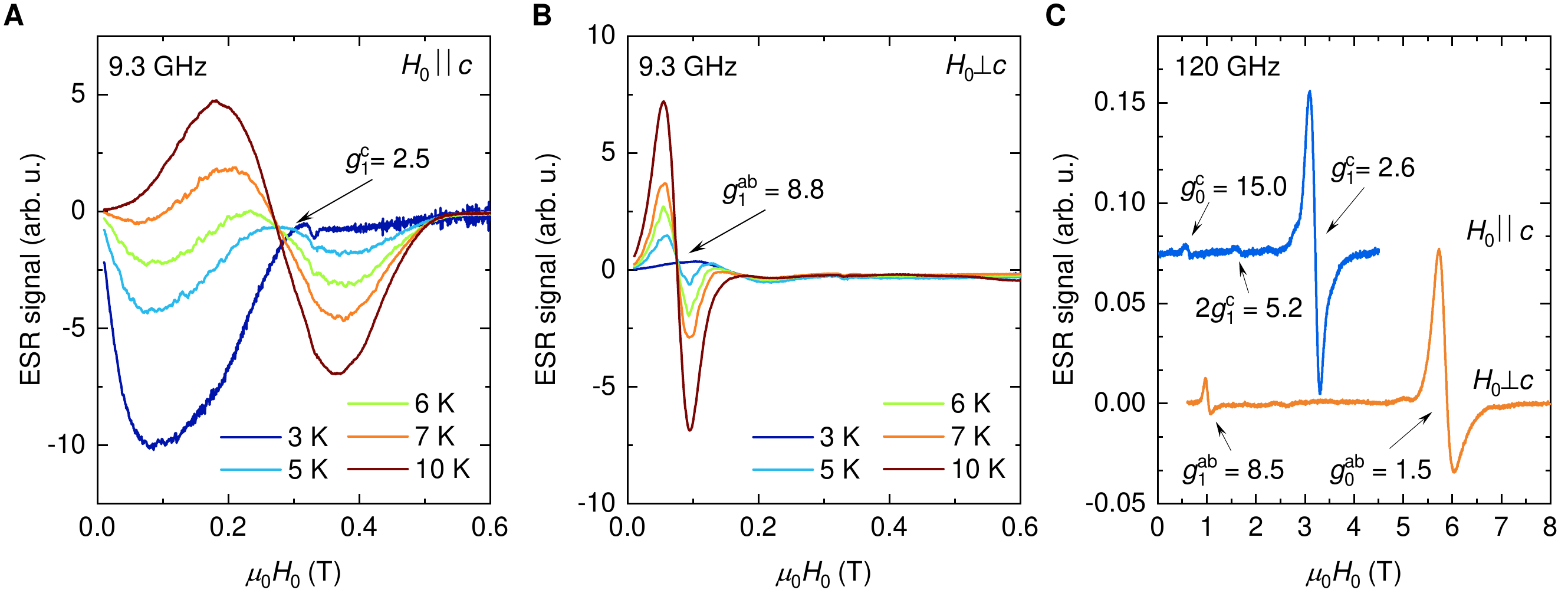}
\caption{{\bf ESR spectroscopy of ErTa$_7$O$_{19}$.} 
(\textbf{A,B}) 
The low-temperature ESR spectra for the magnetic field applied perpendicular to the triangular planes ($H_0{\parallel}c$) and within the triangular planes ($H_0{\perp}c$) at 9.3~GHz.
(\textbf{C})  
The high-frequency ESR spectra recorded at 120~GHz and 70~K for both field directions. 
The spectrum for $H_0{\parallel}c$ is shifted vertically for clarity.
}
\label{figS3}
\end{figure}

The spectra recorded at 9.3~GHz show drastic shape transformation in the temperature range between 3 and 10 K (Fig.~\ref{figS3}A,B) due to CEF effects.
For $H_0{\parallel} c$, the low temperature signal is found at very small field, signaling $g$ factor $g_0^\text{c}>10$. However, due to the large width of this signal, and its vicinity to zero field, exact value of $g_0^\text{c}$ cannot be reliably determined. On the other hand, the signal that dominates at higher temperatures is centered at $g_1^\text{c} = 2.5\pm 0.1$ (Fig.~\ref{figS3}A).
The situation is reversed for $H_0{\perp} c$, where the low-temperature signal is found at higher fields (its exact position again cannot be determined due to the large width of the signal), while the high-temperature signal is centered at lower fields corresponding to the $g$ factor $g_1^\text{ab} = 8.8\pm 0.2$, as shown in Fig.~\ref{figS3}B.

The $g$ factors of the signal from the first excited Kramers doublet were confirmed to be at $g_1^\text{c} = 2.6\pm 0.1$ and $g_1^\text{ab} = 8.5\pm 0.2$ via the high-frequency measurement, which, in addition, revealed the $g$ factors of the ground-state Kramers doublet (Fig.~\ref{figS3}C).
These are $g_0^\text{c}=15.0\pm 0.1$ and $g_0^\text{ab} = 1.5\pm 0.1$.
The positions of all measured ESR signals correspond very well to $g$ factors  $g_0^\text{c}=15.1\pm 0.2$ and $g_0^\text{ab}=1.4\pm 0.1$ determined in the mean-field fit of the low-temperature bulk magnetization, as well as values $g_0^\text{c}=14.9\pm 0.2$, $g_0^\text{ab}=1.2\pm 0.2$ and $g_1^\text{c}=2.5\pm 0.1$, $g_1^\text{ab}=8.8\pm 0.2$ obtained from CEF modeling.

\subsubsection*{Determination of magnetic interactions}
The dipolar Hamiltonian on the three-dimensional lattice of ErTa$_7$O$_{19}$ is  
\begin{equation}
\mathcal{H}_\text{d} = \sum_{(ij)} \sum_{\alpha\beta} J_{\text{d},ij}^{\alpha\beta} S_i^\alpha S_j^\beta ,
\label{eq_dip_hamiltonian}
\end{equation}
where the sum runs over all spin pairs $(ij)$ with $i \neq j$, $\alpha, \beta = \text{x, y, z}$, and $J_{\text{d},ij}$ is a symmetric dipolar coupling tensor between spins $i$ and $j$.
For a uniaxial $g$ tensor with eigenvalues $g_\text{c}$ and $g_\text{a} = g_\text{b} = g_\text{ab}$ along and perpendicular to the $c$ axis, respectively, this tensor is given by
\begin{equation}
J_{\text{d},ij}^{\alpha\beta} = \frac{\mu_0 \mu_\text{B}^2 g_\alpha g_\beta}{4 \pi r_{ij}^3} \left(1 - \frac{3 r_{ij}^\alpha r_{ij}^\beta}{r_{ij}^2} \right) ,
\label{eq_dip_coupling_tensor}
\end{equation}
where $\mathbf{r}_{ij}$ is the displacement vector from spin $i$ to spin $j$.
Since in ErTa$_7$O$_{19}$ $g_0^\text{c} = 15.0 \gg g_0^\text{ab} = 1.5$, the most dominant dipolar couplings by far are between spin components along the $c$ axis, i.e., $J_{\text{d},ij}^\text{zz} \gg J_{\text{d},ij}^{\alpha\beta}$ for $\alpha, \beta \neq \text{z}$. 
The $nn$ and $nnn$ dipolar couplings within the triangular planes, $J_\text{d1}$ and $J_\text{d2}$, respectively, as well as in neighboring planes, $J_\text{d1}'$ and $J_\text{d2}'$, respectively, are illustrated in Fig.~\ref{fig1}A and their components summarized in Table~\ref{tabdipol}.

\begin{table}[t]
\centering
\caption{
{\bf Magnetic interactions in ErTa$_7$O$_{19}$.}
The components of the dipolar interaction between nearest neighbors ($J_{\text{d}1}^{\alpha}$) and next-nearest neighbors ($J_{\text{d}2}^{\alpha}$) within triangular planes in ErTa$_7$O$_{19}$, as well as between nearest neighbors ($J_{\text{d}1}'^{\alpha}$) and next-nearest neighbors ($J_{\text{d}2}'^{\alpha}$) in consecutive planes (all in mK per $k_\text{B}$) for a given displacement vector expressed in fractional coordinates. 
These interactions and the corresponding microscopic contribution to the dipolar Weiss temperature $\theta^\alpha_\text{CW,d}$ (in mK) are calculated exactly from experimental lattice parameters and $g$ factors. 
The diagonal components of the nearest-neighbor exchange interaction $J_\text{e}^{\alpha}$, on the other hand, are calculated via Equation~\ref{eq_ex_weiss} from the exchange Weiss temperature $\theta^\alpha_\text{CW,e}$ derived from Equation~\ref{eq_weiss}.\\
\label{tabdipol}}
\begin{tabular}{c|c|c c c}
 \hline
 Interaction & Displacement & $\alpha=\text{x}$ & $\alpha=\text{y}$ & $\alpha=\text{z}$ \\
               \hline
$J_{\text{d}1}^{\alpha}$ ($\text{mK}\cdot k_\text{B}$) & $(1,0,0)$ & -11.8 & 5.9 & 587     \\
$J_{\text{d}2}^{\alpha}$ ($\text{mK}\cdot k_\text{B}$) & $(1,1,0)$ & -1.4 & 0.3  & 113     \\
$J_{\text{d}1}'^{\alpha}$ ($\text{mK}\cdot k_\text{B}$) & $(0,0,1)$ & 1.4 & 1.4 & -285     \\
$J_{\text{d}2}'^{\alpha}$ ($\text{mK}\cdot k_\text{B}$) & $(1,0,1)$ & 0.2 & 0.9 & -101 \\
$\theta^\alpha_{\text{CW,d}_\text{mic}}$ (mK) & / & 3.7 & 3.7 & -730 \\
\hline
$\theta^\alpha_\text{CW,e}$ (mK) & / & -78(4) & -78(4) & -53(6)\\
$J_\text{e}^{\alpha}$ ($\text{mK}\cdot k_\text{B}$) & $(1,0,0)$ & 52(3) & 52(3) & 35(4)\\
 \hline
\end{tabular}
\end{table}

\begin{figure}[h!]
\centering
\includegraphics[trim = 0mm 0mm 0mm 0mm, clip, width=1\linewidth]{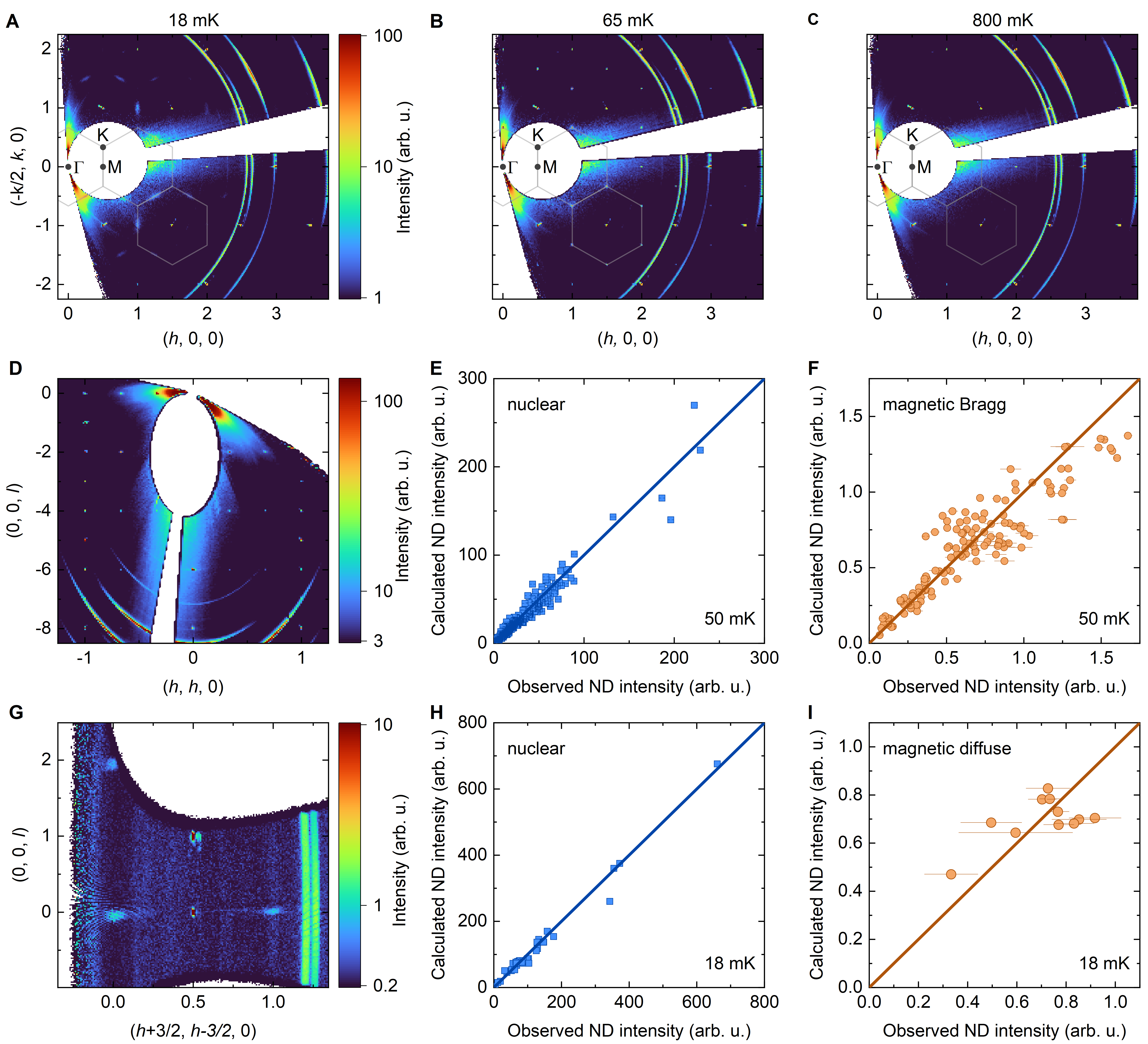}
\caption{{\bf  Single-crystal neutron diffraction above and below the N\'eel temperature $T_\text{N} = 100$~mK.} 
(\textbf{A--C})
Structural Bragg peaks at $\Gamma$ points, magnetic Bragg peaks at K points [magnetic vector ${\bf q}_\text{K}=(1/3,1/3,0)$], and magnetic diffuse signal at M points [magnetic vector ${\bf q}_\text{M}=(1/2,0,0)$].
The ring-like features are due to the copper sample holder.
Selected Brillouin zone boundaries are shown by thin solid lines.
(\textbf{D})
The nuclear Bragg peaks and the magnetic Bragg peaks corresponding to ${\bf q}_\text{K}$ at 65 mK in a plane at the position $(k,-k)=(0,0)$.
(\textbf{E,F}) 
Comparison of calculated and observed intensities of 465 nuclear peaks ($R_\text{F}=8.3$) and 130 magnetic peaks ($R_\text{F}=8.7$) at 50~mK.
(\textbf{G}) 
Nuclear Bragg peaks and diffuse magnetic peaks corresponding to ${{\bf q}_\text{M}}$ at 18 mK in a plane at the position $(k,-k)=(3/2,-3/2)$.
(\textbf{H,I}) 
Comparison of calculated and observed intensities of 38 nuclear peaks ($R_\text{F}=4.9$) and 11 diffuse magnetic peaks ($R_\text{F}=8.4$) at 18~mK.
}
\label{figS3ND}
\end{figure} 

The microscopic interactions transform into Weiss temperatures in the molecular-field picture.
The exchange constants in Equation~\ref{exdip} yield the exchange contribution to the Weiss temperature,
\begin{equation}
\theta^\alpha_\text{CW,e} = -\frac{1}{4k_B} \sum_{(ij)} J_\text{e}^\alpha = -\frac{z}{4k_\text{B}}J_\text{e}^\alpha,
\label{eq_ex_weiss}
\end{equation}
where $z=6$ is the number of nearest neighbors on the triangular lattice.
In a similar manner, the microscopic contribution of the dipolar interactions to the Weiss temperature for a selected direction $\alpha$,
\begin{equation}
\theta^\alpha_{CW,d_{\rm mic}} = -\frac{1}{4k_B} \sum_{(ij)} J_{d,ij}^{\alpha} ,
\label{eq_dip_weiss}
\end{equation}
can be calculated by considering the dipolar couplings between an (arbitrary) spin $i$ and all its neighbors $j$ in a sphere with the radius of $20$ lattice sites centered at spin $i$, which ensures convergence of the sum.
We obtain an antiferromagnetic $\theta^\text{c}_{\text{CW,d}_\text{mic}} = -730$~mK and ferromagnetic $\theta^\text{ab}_{\text{CW,d}_\text{mic}} = 3.7$~mK (Table~\ref{tabdipol}).
It is important to realize that the Weiss temperatures reflect local fields $H_\text{m} + H_0-NM + H_\text{L}$, which are composed of microscopic fields $H_\text{m}$, originating from exchange and dipolar interactions, as well as macroscopic fields, including the applied field $H_0$, demagnetization field $-NM$ and Lorentz field $H_\text{L}=M/3$. 
The latter accounts for cutting out the Lorentz cavity from continuous medium within which the microscopic fields are exactly calculated~\cite{lee1999muon}.
The Lorentz field thus accounts for the dipolar field originating from outside the Lorentz sphere and yields a macroscopic contribution to the dipolar Weiss temperature
\begin{equation}
\theta^\alpha_{\text{CW,d}_\text{mac}} =\frac{C^\alpha}{3}\frac{\rho}{\mathcal{M}},
\label{eq_Lor}
\end{equation}
where $\rho = 8.80$~g\,cm$^{-3}$ and $\mathcal{M} = 1738$~g\,mol$^{-1}$ are the density and molar mass or ErTa$_7$O$_{19}$, respectively, and $C^\alpha$ is the Curie constant along $\alpha$ direction.
From the experimental
\begin{equation}
\theta^\alpha_\text{CW} = \theta^\alpha_\text{CW,e}+\theta^\alpha_{\text{CW,d}_\text{mic}}+\theta^\alpha_{\text{CW,d}_\text{mac}}
\label{eq_weiss}
\end{equation}
we can thus estimate the exchange contributions to the Weiss temperatures $\theta^\text{c}_\text{CW,e} = -57\pm 6$~mK and $\theta^\text{ab}_\text{CW,e} = -78\pm 5$~mK, yielding the $nn$ exchange constants $J_\text{e}^\text{z}/k_\text{B}= 38\pm 4$~mK and $J_\text{e}^\text{xy}/k_\text{B}=52\pm 3$~mK from Equation~\ref{eq_ex_weiss}.

\subsubsection*{Neutron diffraction}
Single crystal ND measurements~\cite{ND} were performed on the WISH time-of-flight diffractometer~\cite{chapon2011wish} at the ISIS Pulsed Neutron and Muon Source of the Rutherford Appleton Laboratory, UK.
A 20-mg crystal was glued with GE Varnish onto a copper sample holder and the measurements were performed both within $(h,k,0)$ and $(h,h,l)$ scattering planes above and below the N\'eel temperature $T_\text{N}=100$~mK (Fig.~\ref{fig3} and Fig.~\ref{figS3ND}).
In each plane, the data were collected for several orientations of the crystal with respect to the incoming neutron beam, in order to capture as many structural and magnetic peaks as possible and to be able to model extinction.

\begin{figure}[h!]
\centering
\includegraphics[width=1\linewidth]{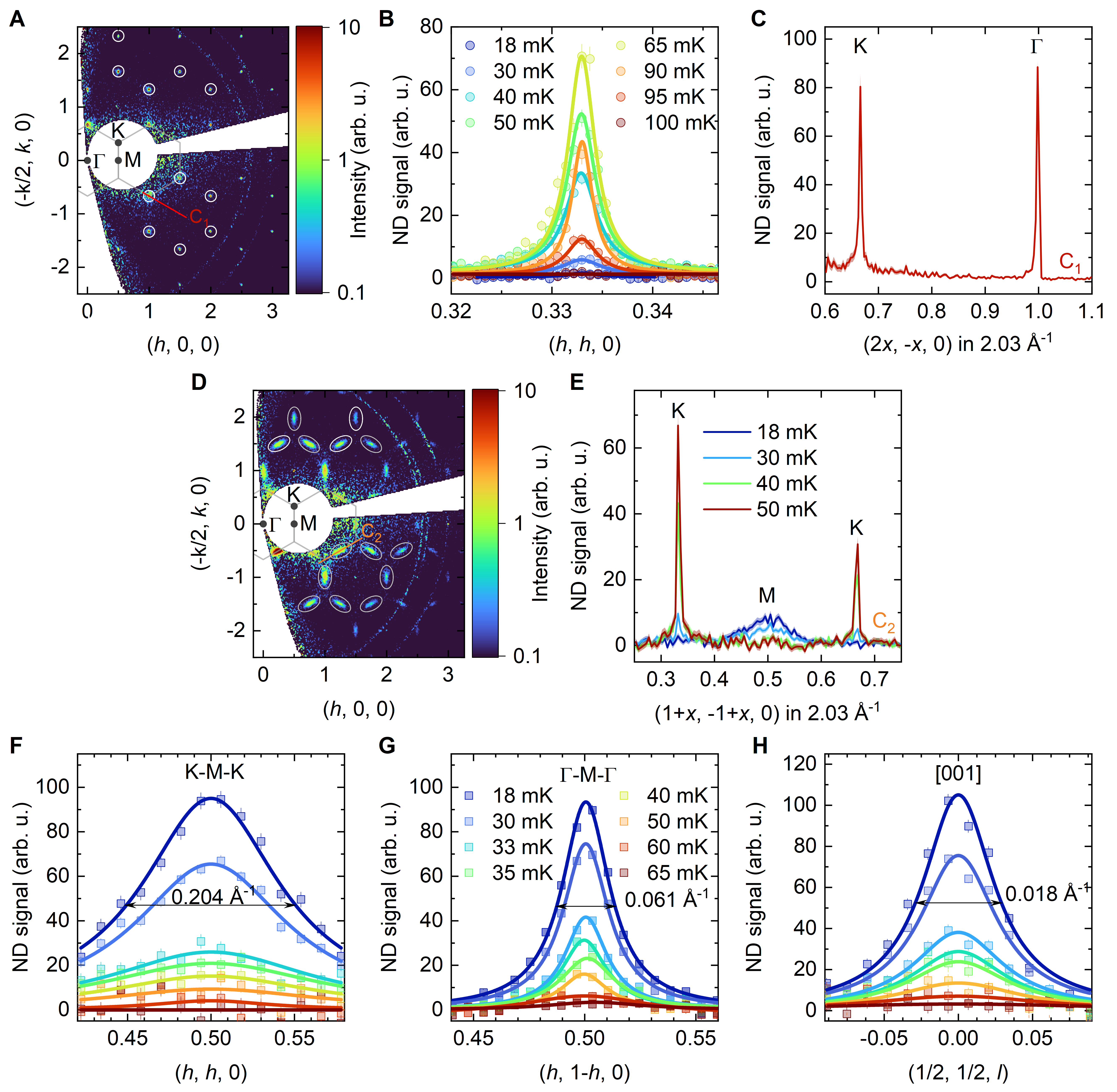}
\caption{{\bf Single-crystal magnetic neutron diffraction in ErTa$_7$O$_{19}$.} 
(\textbf{A})
Magnetic Bragg peaks at 65~mK highlighted by gray circles used to extract the temperature dependence of the ordered magnetic moment.
(\textbf{B}) 
The corresponding Lorentzian fits (solid lines) of the averaged signal (symbols).
(\textbf{C}) 
A cut along ${\rm C_1}=(2x,-x,0)$, as indicated in ({\bf A}), of the 65-mK dataset (Fig.~\ref{figS3ND}B) covering the magnetic Bragg peak at $(4/3,-2/3,0)$ and the structural Bragg peak at $(2,-1,0)$.
(\textbf{D}) 
The diffuse pattern at 18~mK with highlighted regions (gray ellipses) used to obtain averaged signal in various cut directions; K-M-K, $\Gamma$-M-$\Gamma$ and $[001]$. 
(\textbf{E}) 
The temperature dependence of the signal for a cut along ${\rm C_2}=(1+x,-1+x,0)$, as indicated in 
({\bf D}).
(\textbf{F--H}) 
The temperature dependence of the diffuse signal (symbols) along the K-M-K, $\Gamma$-M-$\Gamma$ and $[001]$ directions, with corresponding Lorentzian fits (solid lines) and arrows indicating the width $\Delta Q_\text{FWHM}$ of the data at 18~mK.
}
\label{figS5}
\end{figure}

Magnetic Bragg peaks appear below $T_N$ in the $\text{LR}_\text{K}$ state only at even $l$ (Fig.~\ref{figS3ND}D,G), revealing FM order between planes, as the unit cell contains two planes. 
Such order is in agreement with the FM interlayer coupling originating from the dipolar interactions.
The width of the magnetic and structural Bragg peaks is very similar and is resolution-limited (Fig.~\ref{figS5}C).
It sets the lower bound of the correlation length to about 400~\AA~and demonstrates high quality of the single crystals.
The corresponding peak intensities were calculated at 50~mK using the skew-integration method within \textsc{Mantid} software~\cite{arnold2014mantid}.
454 structural peaks within the ranges $-4\leq h\leq 3$, $-3\leq k\leq 7$ and $-24\leq l\leq 11$ were used for structural refinement within the {\sc Fullprof} suite~\cite{rodriguez1993recent} (Fig.~\ref{figS3ND}E and Table~\ref{struct}). 
The obtained scale was then used for magnetic refinement, where the intensities of 130 magnetic peaks at K points, corresponding to the magnetic propagation vector ${\bf q}_\text{K}=(1/3,1/3,0)$, were taken into account.
The modeling was performed by considering irreducible representations (irreps) of the little group for ${\bf q}_\text{K}$.
The representation analysis yields six one-dimensional irreps $K_i$, each appearing once in the little group so that each comprises a single basis vector that connects the two crystallographically equivalent Er$^{3+}$ magnetic ions within the unit cell (Table~\ref{tab-irreps-basvec-K}).
Out of the six possible basis vectors for the $\text{LR}_\text{K}$ state, the refinement of the magnetic reflections yields a good result only for the basis vector $\psi_{3}^{{\bf q}_\text{K}}$ (Fig.~\ref{figS3ND}F).
This corresponds to the $\frac{\rm U}{2}\frac{\rm U}{2}{\rm D}$ Ising-type order with the amplitude $m_\text{z} = 5.2\pm 0.1\mu_\text{B}$ for the out-of-plane component. 
Adding the transverse component $\psi_{2}^{{\bf q}_\text{K}}+\psi_{6}^{{\bf q}_\text{K}}$, corresponding to the in-plane order of the spin supersolid, does not improve the quality of the fit and thus could not be determined from our experiment.
The temperature dependence of the size of the ordered magnetic moment (Fig.~\ref{fig3}D) was obtained from the temperature dependence of the intensity of 11 selected magnetic Bragg peaks (Fig.~\ref{figS5}A), scaling with the square of the magnetic moment. 

A similar approach was used to characterize the $\text{SR}_\text{M}$ state at 18~mK reflected in diffuse magnetic peaks at M points (Fig.~\ref{figS3ND}G).
Due to the large width of the magnetic peaks, the intensity of only the 11 most intense peaks was calculated using the MD-integration method within \textsc{Mantid} software~\cite{arnold2014mantid}.
The integration was performed within the radius of 0.20~\AA$^{-1}$~around each peak, which was found to be large enough to integrate the whole signal in all three dimensions, while the background was determined in the range between 0.22 and 0.24~\AA$^{-1}$.
The scale was determined from the intensity of 38 nuclear Bragg peaks from the same dataset via the same method (Fog.~\ref{figS3ND}H).
The magnetic refinement was performed by considering irreps of the little group for the magnetic propagation vector ${{\bf q}_\text{M}=}(1/2,0,0)$.
The representation analysis yields four real one-dimensional irreps $M_i$, with representations $M_1$ and $M_2$ appearing once and representations $M_3$ and $M_4$ appearing twice in the little group.
The corresponding six basis vectors connecting the two crystallographically equivalent Er$^{3+}$ magnetic ions within the unit cell are summarized in Table~\ref{tab-irreps-basvec-K}.
The refinement of the diffuse magnetic peaks (Fig.~\ref{figS3ND}I) reveals that the short-ranged magnetic correlations are in agreement with the basis vector $\psi_3^{{\bf q}_\text{M},2}$ (irrep $M_3$), corresponding to stripes of magnetic moments of $3.1\pm 0.1\,\mu_\text{B}$ pointing perpendicular to the triangular planes.

\begin{table}[t]
 \centering
 \caption{
{\bf Irreps at K and M points.}
 Basis vectors $\psi_i^{{\bf q}_\text{K}}$ of the irreps $K_i$ ($i$\,=\,1--6) for magnetic vector ${\bf q}_\text{K}=(1/3,1/3,0)$ and $\psi_i^{{\bf q}_\text{M},j}$ of the irreps $M_i$, where $i$\,=\,1--4 and $j$\,=\,1 (1,\,2) for $M_1$ and $M_2$ (for $M_3$ and $M_4$) for magnetic vector ${\bf q}_\text{M}=(1/2,0,0)$ for the space group $P\bar{6}c2$ (No.~188) appearing in the magnetic representation for the magnetic Er site $(1/3,2/3,0)$ at the 2c Wyckoff position.
 The listed basis vectors are scaled to yield normalized magnetic moments in the complex plane with $m_\text{x}$, $m_\text{y}$, and $m_\text{z}$ pointing along the crystallographic $a$, $b$, and $c$ axes, respectively.\\}
   \begin{tabular}{ c|   c|   c c c |  c c c  }
   \hline
irrep & Basis & \multicolumn{3}{c|}{Atom 1} & \multicolumn{3}{c}{Atom 2} \\
      & vector & $m_\text{x}$ & $m_\text{y}$ & $m_\text{z}$ & $m_\text{x}$ & $m_\text{y}$ & $m_\text{z}$ \\
\hline      
$K_1$ & $\psi_1^{{\bf q}_\text{K}}$ & $\sqrt{3}/2-i/2$ & $-i$ & 0 & $-\sqrt{3}/2+i/2$ &  $i$ &    0    \\
$K_2$ & $\psi_2^{{\bf q}_\text{K}}$ & $\sqrt{3}/2-i/2$ & $-i$ & 0 &  $\sqrt{3}/2-i/2$ & $-i$ &    0 \\
$K_3$ & $\psi_3^{{\bf q}_\text{K}}$ &      0           & 0    & 1 &    0              &   0  &    1  \\
$K_4$ & $\psi_4^{{\bf q}_\text{K}}$ &      0           & 0    & 1 &    0              &   0  & $-1$ \\
$K_5$ & $\psi_5^{{\bf q}_\text{K}}$ & $\sqrt{3}/2+i/2$ &  $i$ & 0 & $-\sqrt{3}/2-i/2$ & $-i$ &    0 \\
$K_6$ & $\psi_6^{{\bf q}_\text{K}}$ & $\sqrt{3}/2+i/2$ &  $i$ & 0 &  $\sqrt{3}/2+i/2$ &  $i$ &    0 \\
\hline
$M_1$ & $\psi_1^{{\bf q}_\text{M}}$ & 1 & 1/2  & 0 & $-1$ &  $-1/2$ & 0    \\
$M_2$ & $\psi_2^{{\bf q}_\textbf{}}$ & 1 & 1/2  & 0 &   1  &     1/2 & 0    \\
$M_3$ & $\psi_3^{{\bf q}_\textbf{},1}$ & 0 & $-1$ & 0 &   0  &       1 & 0    \\
$M_3$ & $\psi_3^{{\bf q}_\textbf{},2}$ & 0 &   0  & 1 &   0  &       0 & 1    \\
$M_4$ & $\psi_4^{{\bf q}_\textbf{},1}$ & 0 & $-1$ & 0 &   0  &    $-1$ & 0    \\
$M_4$ & $\psi_4^{{\bf q}_\textbf{},2}$ & 0 &   0  & 1 &   0  &       0 & $-1$ \\
\hline
 \end{tabular}
 \label{tab-irreps-basvec-K}
\end{table}

The correlation lengths in the $\text{SR}_\text{M}$ state were determined from the averaged spin structure factor along the edge of the Brillouin zone (K-M-K direction) and in the perpendicular directions $\Gamma$-M-$\Gamma$ and $[001]$. 
14 equivalent cuts of the ND map (Fig.~\ref{figS5}D) were averaged to improve statistics, and a Lorentzian curve was fitted to the data.
The widths of the signal at the K and M points were then used to determine the orientation-dependent spin correlation length $\xi$ (Fig.~\ref{fig3}E), using the relation~\cite{zhu2005modern}
\begin{equation}
    \xi = \frac{2}{\Delta Q_\text{FWHM}},
\label{ksi}
\end{equation}
where $\Delta Q_\text{FWHM}$ corresponds to the full width of the signal at half maximum.

\subsubsection*{Measurements and DFT calculations of specific heat}
Specific heat $c(T)$ of a 0.34-mg single crystal was measured with a dilution-refrigerator setup attached to the DynaCool Quantum Design PPMS instrument in the temperature range between 65~mK and 4~K.
The measurements were performed in zero magnetic field as well as in fields up to 0.25~T applied in the direction perpendicular to the triangular planes (Fig.~\ref{figS6}D).
At 4~K the measured specific heat in zero applied field is very similar to the contribution of the excited Kramers doublets, as derived from the CEF model (Equation~\ref{eqCEF}) and shown in Fig.~\ref{figS6}E.
This reveals that any phonon contribution is still marginal at this temperature, while the entropy of the ground state Kramers doublet is already fully released.

The nuclear contribution to the specific heat $c_\text{n}$ was estimated using \textit{ab initio} density functional theory (DFT) calculations. 
In these calculations, the electric field gradient (EFG) tensors at the positions of the nuclei were calculated using the \textsc{Castep} plane-wave DFT code~\cite{clark2005first} with ultrasoft pseudopotentials and the PBE exchange--correlation functional~\cite{perdew1996generalized} with additional Hubbard repulsion~\cite{anisimov1997first}. 
Effective on-site Hubbard repulsions of $U_\text{eff} = U - J_\text{H} = 1.0$ and $8.0$~eV, where $U$ is the bare Hubbard repulsion and $J_\text{H}$ is Hund's coupling, were chosen for the Ta $5d$ and Er $4f$ orbitals, respectively, consistently with typical Hubbard repulsion strengths in Sn$_2M$TaO$_6$ ($M$ = Mn, Fe)~\cite{rahmani2022theoretical} and Ba$_2$ErReO$_6$~\cite{haid2021predictive}, respectively.
A 2200~eV plane-wave energy cut-off and $3 \times 3 \times 1$ Monkhorst--Pack grid~\cite{monkhorst1976special} reciprocal-space sampling was chosen to achieve numerical convergence. 
The atomic positions were converged using DFT geometry optimization to within a tight 5~meV\,\AA$^{-1}$ force tolerance on the nuclei, while the self-consistent field DFT loop was converged to within a tight total energy tolerance of 0.1~neV per atom for geometry optimization tasks and 30~$\mu$eV per atom for EFG tensor calculations. 

The EFG tensors $V_i$ were calculated at crystallographic Er, Ta1 and Ta2 sites, with the calculations yielding the largest eigenvalues $q_{0,i}^\text{Er}=1014$~V\,\AA$^{-2}$, $q_{0,i}^\text{Ta1}=52.4$~V\,\AA$^{-2}$, $q_{0,i}^\text{Ta2}=290$~V\,\AA$^{-2}$ and asymmetry parameters $\eta^\text{Er}=\eta^\text{Ta1}=0$, $\eta^\text{Ta2}=0.092$.
These values were used to extract zero-field nuclear energy levels $\{E_{i,j,k}\}$ for a nuclear isotope $j$ at site $i$, where $k = 1, \ldots, 2 I_j+1$ and $I_j$ is the nuclear spin of isotope $j$, by diagonalizing the quadrupolar nuclear Hamiltonian~\cite{gomilsek2023many,ashbrook2006structural,slichter1990principles} 
\begin{equation}
\mathcal{H}_{\text{q},ij} = \frac{e_0 Q_j}{2 I_j (2 I_j - 1)} \mathbf{I}_j \cdot (V_i \mathbf{I}_j) ,
\label{eq_ham_quad}
\end{equation}
where $e_0$ is the elementary charge, $\mathbf{I}_j$ is the vector spin operator of each isotope and $Q_j$ is its nuclear quadrupole moment. 
From the obtained nuclear energy levels, the total nuclear specific heat $c_\text{n}$ per mole of  Er$^{3+}$ magnetic ions was calculated as 
\begin{equation}
c_\text{n} = \frac{R}{N_\mathrm{Er}} \sum_{i,j} N_i f_j \frac{\left\langle E^2\right\rangle_{ij} - \left\langle E\right\rangle_{ij}^2}{(k_\text{B} T)^2}   
\label{eq_spec_heat_quad}
\end{equation}
where $N_i$ is the number of atoms in the unit cell corresponding to the crystallographic site $i$, $f_j$ is the natural abundance of the nuclear isotope $j$, $Z_{ij} = \sum_k e^{-E_{ij,k}/(k_\text{B} T)}$ is the nuclear partition function of isotope $j$ at site $i$, and $\left\langle E\right\rangle ^n_{ij} = \left[ \sum_k E_{ij,k}^n e^{-E_{ij,k}/(k_\text{B} T)} \right] / Z_{ij}$ is the thermal average of the $n$-th power of the nuclear energy of isotope $j$ at site $i$. 
Here, the dependence of nuclear energy levels on the applied field was neglected since the nuclear gyromagnetic ratios are much smaller than that of electrons. 
The contribution of $^{181}$Ta nuclei to $c_\text{n}(T)$ is dominant, while the contribution of $^{167}$Er also represents a non-negligible correction at temperatures above $\sim$30~mK (Fig.~\ref{figS6}E).
The contribution of $^{17}$O nuclei is ${\sim}4$ orders of magnitude smaller  due to the low natural abundance and small nuclear quadrupolar moment of $^{17}$O.

The magnetic contribution to the specific heat shown in Fig~\ref{fig2}E was then obtained as $c_\text{m}= c-c_\text{n}$.
This was in turn used to calculate the temperature dependence of the entropy (Fig.~\ref{fig2}F) as $S(T)=R\,\text{ln}(2)-\int_{T}^{T_\text{c2}}c_\text{m}dT/T$.
We set $T_\text{c2} =2$~K, because at this temperature the measured specific heat exhibits a minimum and quickly approaches the contribution of the excited Kramers doublet above this temperature (Fig.~\ref{figS6}E), revealing a fully released entropy of the ground state Kramers doublet, as expected for $T_\text{c2} \gg J_1$.
The validity of our estimate of $c_\text{n}$ is confirmed by measurements in 0.1~T.
For this field we find $c_\text{n}(T_\text{c1})\simeq c(T_\text{c1})$ (Fig.~\ref{figS6}E) for the lowest temperature we could reach, $T_\text{c1}=65$~mK$\ll T_\text{N}'=143$~mK, in agreement with the gapped UUD state in the 1/3 magnetization plateau.
Moreover, the extracted entropy release between $T_\text{c1}$ and $T_\text{c2}$ matches the theoretical value of $R\ln(2)$ within a few percent (Fig.~\ref{fig2}F).

\subsubsection*{FTLM calculations}
Finite temperature Lanczos method (FTLM) calculations~\cite{jaklic00finite} were performed on a high-symmetry 2D triangular lattice with $N=36$ sites, containing both relevant wave vectors (${\bf q}_\text{K}$ and ${\bf q}_\text{M}$).
The model described in the main text with finite $\alpha$ requires full quantum many-body calculations with $N_\text{st} \sim 10^{10}$ basis states~\cite{ulaga2024finite}, which are quite delicate in finite-size calculations.
We, therefore, used a recently improved variant of the FTLM~\cite{morita2020}, which eliminates most of the sampling requirements at the cost of storing one additional wave-function. 
We calculated the matrix elements of the spin structure factor in the Lanczos basis, further reducing memory costs. 
The central quantity evaluated within FTLM is the grand-canonical sum 
$Z(T)={\rm Tr} \left\{ {\rm exp}\left[-(\mathcal{H}-E_0)/(k_\text{B} T) \right] \right\}$, where $E_0$ is the ground state energy.
$Z(T)$ was then used to evaluate the entropy
$S(T)=R\left[{\rm ln}Z+ \left(\left\langle\mathcal{H} \right\rangle - E_0 \right)/T \right]/N$ and the corresponding specific heat $c_\text{m}(T)=T({\rm d}S/{\rm d}T)$. 
The static spin structure factor was calculated as
$S_{\bf q}(T)=\sum_{i,j}{\rm exp}\left[i{\bf q}({\bf r}_i-{\bf r}_j) \right] \langle S_i^\text{z} S_j^\text{z} \rangle/N$, where the sums ran over all sites.
Parts of the calculations were performed with the XDiag library~\cite{wietek2025xdiag}.

\subsubsection*{Mean-field modeling of 3D ordering}
The 3D nature of the ordering at the magnetic propagation vector ${\bf q}_\text{K}=(1/3,1/3,0)$ below $T_\text{N} = 100$~mK, being due to sizable FM interlayer interactions of dipolar origin, was investigated within the mean-field (MF) approach.  
Within this approach, an effective local field $h^\text{MF}_i$ emerging from neighboring layers is added to the single-layer Hamiltonian, assuming equal magnetization $m_{{\bf q}_\text{K}} \ne 0$ within each layer,
\begin{eqnarray}
h^\text{MF}_i &=& - \sum_{j} J'_{\text{d},i j} \langle S^\text{z}_{j} \rangle = \mathrm{e}^{- i {{\bf q}_\text{K}}\cdot {\bf R}_i } 
h^\text{MF}_{{\bf q}_\text{K}},  \nonumber \\
h^\text{MF}_{{\bf q}_\text{K}} &=& - J^\text{MF}_{{\bf q}_\text{K}} m_{{\bf q}_\text{K}} ,\; J^\text{MF}_{{\bf q}_\text{K}}  = 
\sum_{j} J'_{\text{d},i j} \mathrm{e}^{- i {{\bf q}_\text{K}}\cdot {\bf R}_{ij} },
\end{eqnarray}
where the sums run over all sites in neighboring layers. 
The induced magnetization within a layer is given by the corresponding susceptibility
\begin{equation}
m_{{\bf q}_\text{K}} \sim  \chi^0_{{\bf q}_\text{K}}  h^\text{MF}_{{\bf q}_\text{K}} \sim - \frac{1}{k_\text{B}T} S_{{\bf q}_\text{K}}(T) J^\text{MF}_{{\bf q}_\text{K}} m_{{\bf q}_\text{K}},
\end{equation}
where we assume the standard relation $\chi^0_{{\bf q}_\text{K}} \sim S_{{\bf q}_\text{K}}/(k_\text{B}T)$, valid because relevant 
dynamic modes satisfy the condition $\hbar\omega <k_\text{B}T$ for long-range order. 
This finally leads to the estimate of $k_\text{B}T_\text{N} \sim -S_{{{\bf q}_\text{K}}}(T_\text{N}) J^\text{MF}_{{\bf q}_\text{K}}$. 
Exact calculation of the effective interlayer dipolar interaction in ErTa$_7$O$_{19}$ gives 
$J^\text{MF}_{{\bf q}_\text{K}}/J_{1} = -0.20$, which together with the rather general behavior of $S_{{\bf q}_\text{K}}$ at high-temperatures (Fig.~\ref{fig4}C) yields $T_\text{N} \sim 0.3 J_{1}/k_\text{B} \sim 0.2$~K.
The overestimate compared to the experimental ordering temperature $T_\text{N} = 100$~mK is generally expected within MFA. 
Due to the abrupt decrease of $S_{{\bf q}_\text{K}}(T)$ at the crossover to the M phase at $T_1$ and the fact that $J^\text{MF}_{{\bf q}_\text{M}}/J_{1} = -0.31$ is similar to $J^\text{MF}_{{\bf q}_\text{K}}/J_{1}$, the transition to a long-range ordered stripe phase should appear around $T_1$.

%



\end{document}